\documentclass[twocolumn,amssymb,floatfix,aps]{revtex4}

\usepackage{amssymb,amsmath}
\usepackage{graphicx}
\usepackage{bm}
\usepackage{epsfig}
\usepackage[section]{placeins}
\newcommand{\bea}{\begin{eqnarray}}
\newcommand{\eea}{\end{eqnarray}}
\def\be{\begin{eqnarray}}
\def\ee{\end{eqnarray}}

\renewcommand{\cos}[1]{\,{\rm cos}(#1)}
\renewcommand{\sin}[1]{\,{\rm sin}(#1)}
\def\pslash{\not{\hbox{\kern-2pt p}}}
\begin{document}
\preprint{CALT 68-2790 }
\title{
Kinematical variables towards new dynamics  at the LHC}

%

\author{Christopher Rogan}

\affiliation{  \vspace{3mm}
Lauritsen Laboratory of Physics, California Institute of Technology, Pasadena, CA 91125}

\date{January 11, 2010}
\date{\today, first draft January 11, 2010}

\begin{abstract}
  At the LHC, many new physics signatures feature the pair-production
  of massive particles with subsequent direct or cascading decays to
  weakly-interacting particles, such as SUSY scenarios with conserved
  conserved R-parity or $H \to W(\ell\nu)W(\ell\nu)$. We present a set of
  dimension-less variables that can assist the early discovery of
  processes of this type in conjunction with a set of variables with
  mass dimension that will expedite the characterization of these
  processes.
\end{abstract}
\pacs{Pacs numbers}  
\maketitle

\section{Introduction\label{sec:intro}}
The LHC is currently delivering high energy $pp$ collisions and
in the coming months and years it will be exploring the TeV
scale. Many of the previously unobserved processes that experimenters
will be searching for involve high transverse momentum Standard Model
(SM) particles, such as leptons and jets, along with missing
transverse momentum -- a characteristic of a large class of models
with conserved, discrete quantum numbers, resulting in new particles
being produced in pairs and decaying to an even
number of stable, weakly-interacting particles that will escape
detection.

In the past years the development of kinematical variables that assist
the discovery of such processes has been intense and rich
\cite{Randall:2008rw}-\cite{Kim:2009si}.
Here we introduce the variables $M_{R}$ and $M_{R^{*}}$, whose
distributions contain information about the masses of pair-produced
particles and the weakly-interacting particles resulting from their
decays. Additionally, we discuss the dimension-less variables $R$ and
$R^{*}$ that can be used to select events of interest in the presence
of large backgrounds, in a number of different exclusive and inclusive
final states.

\section{$M_R$}\label{sec:RFRAME}

To define the $M_R$ we consider a simple example. We assume two
massive particles, $G_{1}$ and $G_{2}$, are produced through a hard
partonic subprocess in a hadron-hadron collision. Furthermore, we
assume that these two particles have the same mass, $M_{G}$. In the
$G_{1}G_{2}$ rest frame (CM frame) the particles $G_{1}$ and $G_{2}$
have equal and opposite momentum, with four-vectors which we define as
\begin{eqnarray}
  \label{eq:CM}
   p[G_1]&\equiv &p_1= M_{G}\,\gamma_{\mathrm{CM}}\,\{1,\vec{\beta}_{\mathrm{CM}}\} 
\nonumber\\
   p[G_2]&\equiv &p_2=M_{G}\,\gamma_{\mathrm{CM}}\,\{1,-\vec{\beta}_{\mathrm{CM}}\},
   \end{eqnarray}
such that $(p_1+p_2)^{2}=\hat{s}$, where $\hat{s}$ is the usual Mandelstam variable describing the hard partonic subprocess. We further assume that each of the two particles $G_i$ decays as follows: $G_i \to Q_i \chi_i$.  We assume each $Q_i$ is a stable, mass-less particle that will be visible in our detector. Each $\chi_i$ is assumed to be stable, potentially massive (with mass $M_{\chi}$) and weakly interacting such that it  escapes detection. In their respective $G_i$ rest frames, the decay products of each $G_i$ have four-momenta defined as
\begin{eqnarray}
\label{eq:Grest}
   p[Q_i]&\equiv &q_i= \frac{M_{\Delta}}{2}\,\{1,\hat{u}_i\}
\nonumber\\
   p[W_i]&\equiv &\omega_i= \frac{M_{\Delta}}{2}\,\{R_{G\chi},-\hat{u}_i\},
   \end{eqnarray}
where $M_{\Delta}=\frac{M_{G}^{2}-M_{\chi}^{2}}{M_{G}}$, $R_{G\chi} = \frac{M_{G}^{2}+M_{\chi}^{2}}{M_{G}^{2}-M_{\chi}^{2}}$ and each $\hat{u}_i$ is a unit vector. To go from the rest frame of $G_1$ ($G_2$) to the CM frame, $q_1$ and $\omega_1$ ($q_2$ and $\omega_2$) are boosted to a frame traveling at a velocity $\vec{\beta}_{\mathrm{CM}}$ ($-\vec{\beta}_{\mathrm{CM}}$) with respect to the $G_1$ ($G_2$) rest frame.  Finally, we assume that to move from the CM frame to the lab frame, each of the final state particles is boosted to a frame traveling at a velocity $\vec{\beta_L} = ( \vec{\beta}_{T}, \beta_{l} )$, where $\vec{\beta}_{T}$ and $\beta_{l}$ are the transverse and longitudinal components of this boost respectively. The transformations taking the final state particles from their respective $G_i$ rest frames to the lab frame can be schematically described as:
\bea
q_1,\omega_1 \xrightarrow{\vec{\beta}_{\mathrm{CM}}} q^{'}_1,\omega^{'}_1  \xrightarrow{\vec{\beta}_{L}} q^{l}_1,\omega^{l}_1 
\nonumber\\
q_2,\omega_2 \xrightarrow{-\vec{\beta}_{\mathrm{CM}}} q^{'}_2,\omega^{'}_2  \xrightarrow{\vec{\beta}_{L}} q^{l}_2,\omega^{l}_2~,
\eea
where $q_{1}^{l}$, $q_{2}^{l}$, $\omega_{1}^{l}$ and $\omega_{2}^{l}$ are the lab frame four-vectors of $Q_{1}$, $Q_{2}$, $\chi_{1}$ and $\chi_{2}$, respectively.

In practice, $|\vec{\beta}_{T}| \sim p_{T}^{ISR} / \sqrt{\hat{s}} \sim p_{T}^{ISR}/ 2M_{G}$, where $p_{T}^{ISR}$ is the magnitude of the vectorial sum of the transverse momenta of initial state radiation. As a result $|\vec{\beta}_{T}| \ll 1$ for sufficiently large values of $M_{G}$. We will use the approximation $\vec{\beta}_{T} \to 0$ for the remainder of the discussion. 

The relevant experimental observables are the three momenta of the massless final state particles, $Q_{1}$ and $Q_{2}$ and the missing transverse momentum, denoted $\vec{M}$, which is the vectorial sum of the transverse momenta of the particles $\chi_{1}$ and $\chi_{2}$. Even with the aforementioned approximation $\vec{\beta}_{T} \to 0$, it is not possible to fully solve for  each of the kinematical unknowns using the available observables in this example.

However there is an additional, well-motivated approximation that can be made to significantly simplify  the problem. If the mass $M_{G}$ is sufficiently large, relative to the hadron-hadron collider energy $\sqrt{s}$, the particles $G_{1}$ and $G_{2}$ will be mostly produced near the $\sqrt{\hat{s}} \sim 2M_{G}$ threshold, such that $\gamma_{CM} \sim 1$. 

Assuming $\gamma_{CM} = 1$ implies, of course, that $\vec{\beta}_{CM}
= 0$. In this approximation, the CM frame of this hard process is now
also simultaneously the rest frame of the particles $G_{1}$ and
$G_{2}$. The momenta of their decay products is given by Equation
~\ref{eq:Grest}, with $q_{i},\omega_{i} =
q^{'}_{i},\omega^{'}_{i}$. Importantly $|\vec{q}_{1}| = |\vec{q}_{2}| = M_{\Delta}/2$ in this frame.

Hence  we can move from the laboratory frame to the CM frame by finding the longitudinal boost to a reference frame where the {\it magnitude} of the momenta of the objects $Q_{1}$ and $Q_{2}$ are equal. We  denote this reference frame the {\it rough approximation}-frame, or $R$-frame, and the longitudinal boost moving from the lab frame to the $R$-frame as $\beta_{R}$. The momenta of $Q_{1}$ and $Q_{2}$ in the $R$-frame are denoted $\vec{q}^{R}_{1}$ and $\vec{q}^{R}_{2}$, respectively. With the constraint $|\vec{q}^{R}_{1}| = |\vec{q}^{R}_{2}|$, we find that
\begin{equation}
\beta_{R} = \frac{q^{l}_{10}-q^{l}_{20}}{q^{l}_{1z}-q^{l}_{2z}}~.
\end{equation}

Furthermore, we define the $R$-frame mass, $M_{R}$, as
\begin{equation}
M_{R} \equiv 2|\vec{q}^{R}_{1}| = 2|\vec{q}^{R}_{2}| = 2\sqrt{\frac{(q^{l}_{10}q^{l}_{2z}-q^{l}_{20}q^{l}_{1z})^{2}}{(q^{l}_{1z}-q^{l}_{2z})^{2}-(q^{l}_{10}-q^{l}_{20})^{2}}}
\label{eq:MR}
\end{equation}

As $\gamma_{CM} \to 0$, we find that $\vec{q}_{i}^{R} \to \vec{q}_{i}$, $\beta_{R} \to \beta_{l}$ and $M_{R} \to M_{\Delta}$. It should also be noted that the quantity $M_{R}$ is invariant under longitudinal boosts so even if $\gamma_{CM} \ne 1$, $M_{R}$ is independent of the true value of $\beta_{l}$.

\section{The $\gamma_{CM} = 1$ approximation \label{sec:gCM}}

In order to understand how $\gamma_{CM}$ is distributed we consider
the simple model with two scalar particles: $\Phi_{0}$ with zero mass
and $\Phi_{1}$ with mass $M_{G}$. Using the notation of
Sec.~\ref{sec:RFRAME}, we consider contact interaction pair
production of $\Phi_{1}$ through a $\Phi_{0}^{2}\Phi_{1}^{2}$
vertex. The sub-process cross-section is proportional to
\begin{equation}
\hat{\sigma}(\hat{s}) \propto \lambda^{2} \frac{\sqrt{1-4M_{G}^{2}/\hat{s}}}{\hat{s}} \propto  \lambda^{2}\frac{\sqrt{1-1/\gamma_{CM}^{2}}}{\gamma_{CM}^{2}M_{G}^{2}},
\label{eq:shat}
\end{equation}
where $\lambda$ is the dimensionless $\Phi_{0}^{2}\Phi_{1}^{2}$ coupling, which we set to 1. From Eq.~\ref{eq:shat} we observe that $\gamma_{CM} = 1$ is kinematically forbidden, and that the cross-section for the sub-process will decrease asymptotically as $1/\gamma_{CM}^2$. 

Additional suppression of large values of $\gamma_{CM}$ is caused by the parton distribution functions (PDFs) in hadron-hadron collisions. Assuming the two initial state $\Phi_{0}$ particles are partons from colliding protons with momentum fractions $x_{a}$ and $x_{b}$ respectively, and PDFs $f_{1}(x)$ and $f_{2}(x)$  we can write the total cross section as:
\begin{equation}
\frac{d\sigma}{dx_{a}dx_{b}} \propto [f_{1}(x_{a})f_{2}(x_{b}) + a \leftrightarrow b]\hat{\sigma}(\hat{s} = sx_{a}x_{b}),
\end{equation}
where $s$ is the proton-proton CM energy. Changing variables from $x_{b}$ to $\gamma_{CM}$ through the relation $sx_{a}x_{b} = 4\gamma_{CM}^{2}M_{G}^{2}$ and integrating over $x_{a}$ we find that the differential cross-section with respect to $\gamma_{CM}$ is given by:
\bea
\frac{d\sigma}{d\gamma_{CM}} \propto
\frac{\sqrt{1-1/\gamma_{CM}^{2}}}{ s\gamma_{CM}} \times \nonumber\\
\int_{\frac{4\gamma_{CM}^{2}M_{G}^{2}}{s}}^{1}[f_{1}(x_{a})f_{2}(\frac{4\gamma_{CM}^{2}M_{G}^{2}}{sx_{a}}) + a\leftrightarrow b] \frac{dx_{a}}{x_{a}}. 
\label{eq:sigma}
\eea

\begin{figure}[htbp]
\begin{center}
\includegraphics[width=0.4\textwidth]{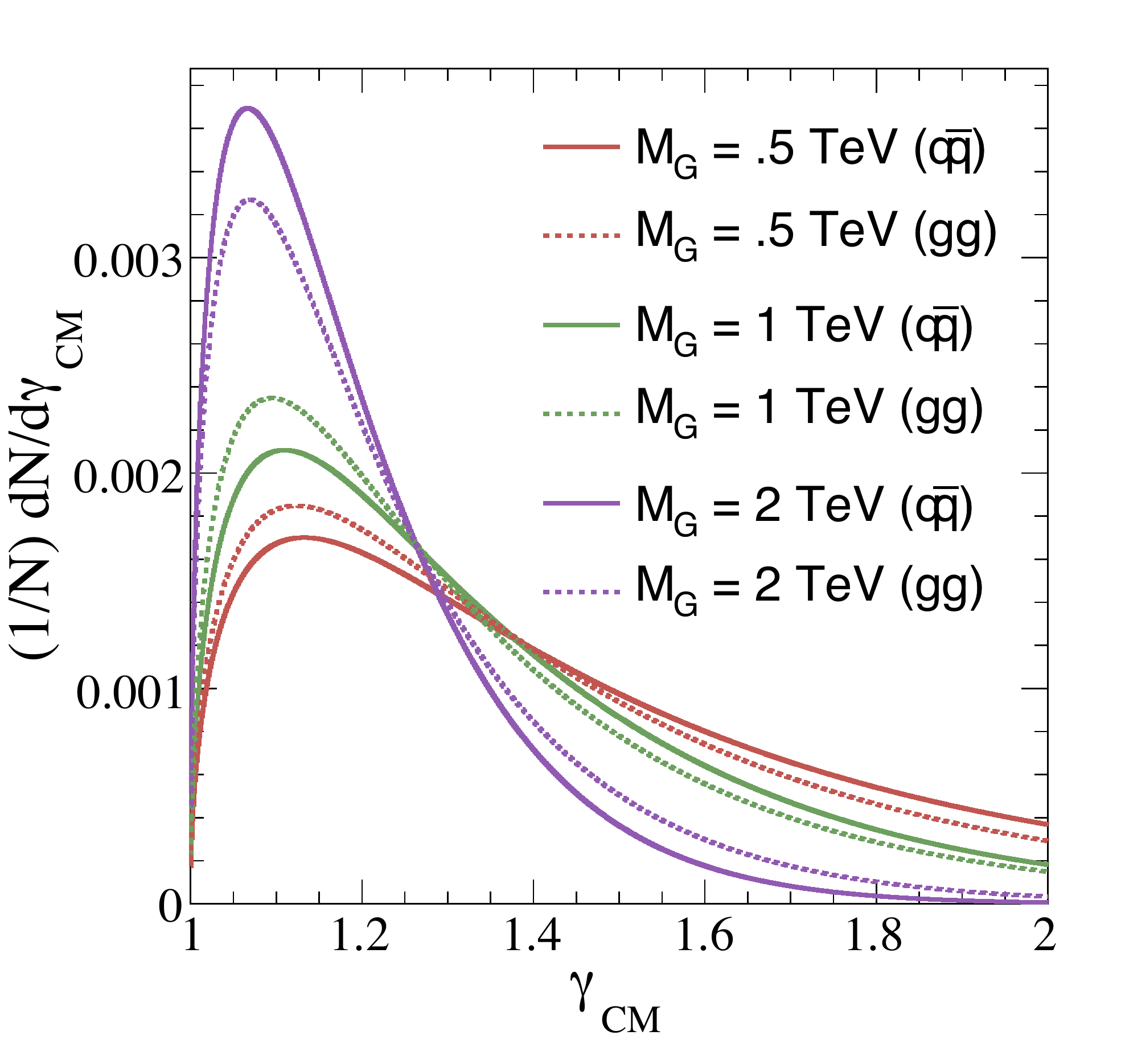}
\caption{Distribution of $\gamma_{CM}$ for $q\bar{q}$-like and $gg$-like production at $\sqrt{s} = 14~\mathrm{TeV}$ for different values of $M_{G}$. 
  \label{fig:gCM}}
\end{center}
\end{figure}

In Fig.~\ref{fig:gCM}  we show the probability distribution function for $\gamma_{CM}$ for $\sqrt{s} = 14$ TeV, where we have numerically integrated Eq.~\ref{eq:sigma} for $q\bar{q}$-like ($u$ and sea quark PDF's) and $gg$-like production.  We use PDF parameterizations of the form $xf_{i}(x) = A_{i} x^{\delta_{i}}(1-x)^{\eta_{i}}(1+\epsilon_{i}\sqrt{x}+\gamma_{i} x) + A_{i}'x^{\delta_{i}'}(1-x)^{\eta_{i}'}$ with NNLO parameters determined from a global PDF fit at $Q^{2} = 1$ GeV$^{2}$~\cite{MSTW}. Larger values of $M_{G}$ result in lower values of $\gamma_{CM}$, with all distributions peaking at approximately $\gamma_{CM} \sim 1.1$ and falling quickly with increasing $\gamma_{CM}$.

The exact dependence of the sub-process cross-section on $\gamma_{CM}$ will vary depending on the nature of the interacting final and initial state particles in the 2 $\to$ 2 process, but the resulting distribution of $\gamma_{CM}$ should be qualitatively similar to the result show in Fig.~\ref{fig:gCM}: $\gamma_{CM}$ exactly equal to 1 is kinematically forbidden, but values of $\gamma_{CM}$ near 1 are preferred to larger values due to the falling sub-process and total cross-sections with increasing $\sqrt{\hat{s}} = \sqrt{sx_{a}x_{b}} \propto \gamma_{CM}$. 

To derive the expression for $M_{R}$ in Eq.~\ref{eq:MR}, we used the approximation $\gamma_{CM}$ precisely equal to 1, and found that $M_{R} \to M_{\Delta}$ as $\gamma_{CM} \to 1$. In order to understand the behavior of $M_{R}$ when $\gamma_{CM} \ne 1$, we return to the example introduced in Sec.~\ref{sec:RFRAME}. Using the same notation, we  again consider the pair production of massive particles $G_{1}$ and $G_{2}$, and continue to use the approximation $\vec{\beta}_{T} \to 0$, this time with $\gamma_{CM}$ not equal to 1. 
\begin{figure}[htbp]
\begin{center}
\includegraphics[width=0.4\textwidth]{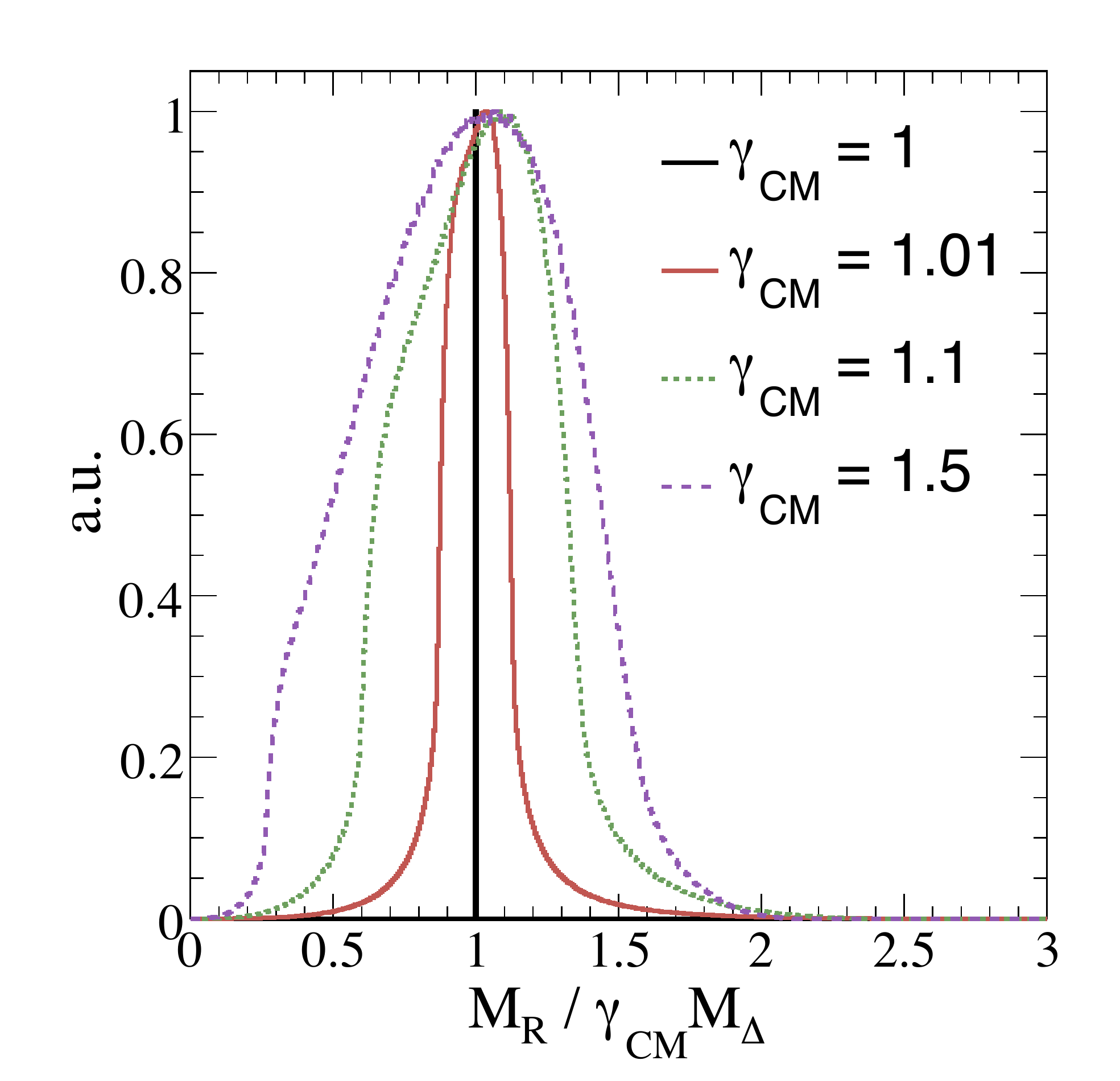}
\caption{Distribution of $M_{R}$, in units of $\gamma_{CM}M_{\Delta}$, for different values of $\gamma_{CM}$. Distributions are normalized such that their maximum value is equal to one.
  \label{fig:MR_can}}
\end{center}
\end{figure}
We recall that the variable $M_{R}$ is invariant under longitudinal boosts, so  its value is independent of the true value of $\beta_{l}$. This implies that for fixed $\gamma_{CM}$, $M_{G}$ and $M_{\chi}$, there are six remaining degrees of freedom, corresponding to the unit vectors $\hat{\beta}_{CM}$, $\hat{u_{1}}$ and $\hat{u_{2}}$. We numerically integrate over these angles, making the assumption that each of these unit vectors are independent of each other with flat probability distribution functions on the unit sphere, as if all the interacting particles are scalars. The resulting distributions of $M_{R}$, for different fixed values of $\gamma_{CM}$, are shown in Fig.~\ref{fig:MR_can}.  We observe that the peak value of $M_{R}$ scales as $\gamma_{CM}M_{\Delta}$, with the width of the $M_{R}$ distribution increasing with $\gamma_{CM}$. Hence, in practice, the distribution of $M_{R}$ will peak near $M_{\Delta}$, even when $\gamma_{CM} \ne 1$. 

\section{Signal and background discrimination: The  razor \label{sec:razor}}
In Sec.~\ref{sec:gCM}, we found that the distribution of the observable $M_{R}$ contains information about the mass difference $M_{\Delta}$, introduced in the example described in Sec.~\ref{sec:RFRAME}. This is a useful property for characterizing this process. 
We further explore whether and when this variable is also useful for selecting events in the presence of background,  establishing the {\it discovery} of this process. This depends on the experimental final state considered and the signal and relevant backgrounds.

We consider massive di-squark production in a generic supersymmetric extension of the Standard Model, where each squark decays directly to the lightest supersymmetric particle (LSP) and a quark, with the LSPs escaping detection. The relevant final state contains two or more jets and missing transverse energy. A particularly challenging background to this final state is QCD multijet production, where non-zero missing transverse energy can result from instrumental backgrounds, jet mis-measurements, finite detector acceptance and non-gaussian tails in the detector response, in addition to the production of neutrinos within jets. 

Using the notation of Sec.~\ref{sec:RFRAME} to describe our signal process, the particles $G_{1}$ and $G_{2}$ are the squarks with mass $M_{\tilde{q}} = M_{G}$, $Q_{1}$ and $Q_{2}$ are the quark jets, and $\chi_{1}$ and $\chi_{2}$ are the LSPs with mass $M_{\chi}$. The final state observables are the two jet four-vectors, $q_{1}^{l}$ and $q_{2}^{l}$, and the missing transverse momentum, $\vec{M}$. 

For the background processes, we consider QCD dijet production. In the dijet rest frame, we express the two jets' four-vectors as:

\begin{eqnarray}
  \label{eq:dijet}
   k_1= \frac{\sqrt{\hat{s}}}{2}\,\{1,\hat{v}\} 
\nonumber\\
   k_2=\frac{\sqrt{\hat{s}}}{2}\,\{1,-\hat{v}\},
   \end{eqnarray}
   where $\sqrt{\hat{s}}$ is the dijet invariant mass and $\hat{v}$ is a unit vector. If we assume that the Lorentz transformation from the dijet rest frame to the laboratory frame is simply a longitudinal boost, $\beta_{l}$,  we find that for this type of event $M_{R} = \sqrt{\hat{s}}$.  Therefore, $M_{R}$ will be distributed as $\sqrt{\hat{s}}$ for this background process, falling steeply, while the signal distribution will peak near $M_{\Delta}$. The question of whether or not we can identify signal events in the presence of this background becomes a question of whether the effective dijet cross-section is sufficiently small for $\sqrt{\hat{s}}$ in the range of the signal peak around $M_{\Delta}$, given a set of  event selection requirements. 

   Traditionally, the requirements that have been used to improve the
   signal to background ratio in such processes is large magnitude of
   the missing transverse energy, $\vec{M}$, as well as large
   transverse momenta of the two jets.  These are highly correlated
   with the observable $M_{R}$. For the background process we are
   considering, the magnitude of the missing transverse energy (from
   mis-measurements of the dijets' transverse momenta or neutrinos
   produced in the jets) is highly correlated with the magnitude of
   the dijets' transverse momentum which, in turn scales with
   $\sqrt{\hat{s}}$. These observables are also highly correlated for
   the type of signal events we are considering since the LSPs'
   transverse momenta will also scale with $M_{\Delta}$, and as a
   result so will the missing transverse energy. A requirement on each
   of these variables is a requirement on the {\it scale} of the
   signal and background events. As a result, for a given integrated
   luminosity, if the background yield in the region $\sqrt{\hat{s}}
   \sim M_{\Delta}$ is prohibitively large relative to the signal
   yield, it is unlikely that additional hard requirements on the
   magnitude of the missing transverse energy or the jet transverse
   momenta will assist the discovery.

   Examining the expression for $M_{R}$ in Eq.~\ref{eq:MR}, we see
   that there is additional kinematical information not yet used. For
   example, $M_{R}$ is independent of the azimuthal angle,
   $\Delta\phi$, between the two final state jets. For the QCD dijet
   background, the jets should be largely back-to-back in the
   transverse plane, with $\Delta\phi$ peaking at $\pi$. On the other
   hand, the two jets in the SUSY signal events result from the decay
   of two separate squarks, implying that their direction in the
   transverse plane is largely independent of each other, apart from
   spin-correlations and effects resulting from $\vec{\beta}_{CM} \ne
   0$. Hence, the distribution of $\Delta\phi$ for signal events will
   be significantly flatter than for the background. Rather than simply
   cutting on the variable $\Delta\phi$, we incorporate this
   information into a new variable  denoted $M_{T}^{R}$.

In this particular final state, we assume that there are two escaping weakly interacting particles, whose four-momenta in the laboratory frame we  denote $\nu^{l}_{1}$ and $\nu^{l}_{2}$, with each particle ``paired'' with an observed jet with four-momenta $q^{l}_{1}$ and $q^{l}_{2}$, respectively. From these four-vectors we define the variable $M_{2G} = \sqrt{(1/2)[(\nu_{1}^{l}+q_{1}^{l})^{2}+(\nu_{2}^{l}+q_{2}^{l})^{2}]}$, which is  equal to $M_{\tilde{q}}$ for signal events. The only constraint we have on the four-vectors $\nu_{i}^{l}$ is that the vectorial sum of their transverse momenta should be equal to the observed missing transverse energy, $\vec{M}$. Setting $(\nu_{i}^{l})^{2} = 0$ and minimizing $M_{2G}$ over $\nu_{1z}^{l}$ and $\nu_{2z}$ yields:
\begin{equation}
\min_{\nu_{iz}} M_{2G} = |\vec{q}_{1T}^{l}||\vec{\nu}_{1T}^{l}| - \vec{q}_{1T}^{l}\cdot \vec{\nu}_{1T}^{l}+|\vec{q}_{2T}^{l}||\vec{\nu}_{2T}^{l}| - \vec{q}_{2T}^{l}\cdot \vec{\nu}_{2T}^{l}
\end{equation}

Motivated by the background we are considering, we assign half of the
measured missing transverse momenta to each escaping particle so that
$\vec{\nu}_{1T}^{l} = \vec{\nu}_{2T}^{l} = \vec{M}/2$ and define
$M_{T}^{R}$ as:
\begin{equation}
M_{T}^{R} = \sqrt{\frac{|\vec{M}|}{2}(|\vec{q}_{1T}^{l}|+|\vec{q}_{2T}^{l}|)-\frac{1}{2}\vec{M}\cdot(\vec{q}_{1T}^{l}+\vec{q}_{2T}^{l})}~.
\label{eq:MT_R}
\end{equation}
The variable $M_{T}^{R}$ also contains information about the scale of the
process we are studying. If we assume that $\gamma_{CM} = 1$ then the
$M_{T}^{R}$ distribution has an endpoint at $M_{\Delta}$ for signal
events. We note that $M_{T}^{R}$ is an additional measurement of the
scale of the process that uses information independent of the
$M_{R}$. Therefore, rather than cutting on $M_{T}^{R}$ we form the
dimension-less '$R$-frame razor', $R$, as the ratio of $M_{T}^{R}$ and
$M_{R}$, such that $R \equiv M_{T}^{R} / M_{R}$. For the signal process,
the distribution of $R$ peaks near 0.5, since this is the ratio of two
measurements of the same scale, $M_{\Delta}$, with an additional
geometrical factor due to the fact that $M_{T}^{R}$ contains only
transverse information. For the QCD dijet background, if $\vec{M} =
0$, then $R$ is 0, for any value of $\sqrt{\hat{s}}$.

As was discussed previously, there are several mechanisms for the
measurement of $\vec{M}$ to be non-zero in QCD dijet events. For
example, one or both jets in the final state could be mis-measured due
to calorimeter non-compensation, uninstrumented regions of the
detector or weakly interacting particles produced within the jets,
causing an imbalance in the event and resulting in non-zero missing
transverse momentum. To evaluate how these possibilities affect the
measured values for $M_{R}$ and $M_{T}^{R}$ in background events, we
return to Eq.~\ref{eq:dijet} which describes the kinematics of the
dijet system in it's CM frame. We now assume that the measured jet
momenta, $q_{i}^{l}$, are scaled relative to their true values, so that
$q_{i}^{l} = f_{i} k_{i}$. Here, we are assuming that the direction of
the two jets is not changed, but rather that only a fraction $f_{i}$
of the jets' momentum is observed, where $f_{i} > 0$. Additionally,
without loss of generality we adopt the convention $f_{1} \ge f_{2}$.

With these mis-measurements, we find that $M_{R}$ takes a value:
\begin{equation}
M_{R} = \sqrt{\frac{4f_{1}^{2}f_{2}^{2}\hat{s}(\hat{v}\cdot \hat{z})^{2}}{(f_{1}+f_{2})^{2}(\hat{v}\cdot \hat{z})^{2}-(f_{1}-f_{2})^{2}}}~,
\label{eq:MR_f}
\end{equation}
independent of the longitudinal boost, $\beta_{l}$, that takes the jets from their CM frame to the laboratory frame. The missing transverse energy can now be non-zero, with $\vec{M} = (f_{2}-f_{1})\vec{k}_{1T}$ and $M_{T}^{R}$ can be expressed as:
\begin{equation}
M_{T}^{R} = \sqrt{(f_{1}-f_{2})f_{1}\frac{\hat{s}(1-(\hat{v}\cdot \hat{z})^{2})}{4}}~.
\label{eq:MT_f}
\end{equation}

From Eq.~\ref{eq:MR_f} we see that these mismeasurements  {\it decrease} the value of $M_{R}$, assuming that $f_{1} \lesssim 1$. Therefore the distribution of $M_{R}$ for the background will not have events promoted to the tail of the distribution due to these types of mis-measurements; instead, these mis-measurements will suppress the background $M_{R}$ distribution. Furthermore, if we require that $R > C$, where $C$ is some cut value, this implies that $C M_{R} < M_{T}^{R}$. To understand the effect of this cut, we  change variables $(\hat{v}\cdot \hat{z})^{2} = \cos{\theta_{1}}^{2}$ and $f_{1} = f_{2}\cos{\theta_{2}}^{2}$. With these substitutions, we re-express the inequality $C M_{R} < M_{T}^{R}$ as:
\bea
16C^{2}\cos{\theta_{1}}^{2}\cos{\theta_{2}}^{4}+\sin{\theta_{1}}^{4}\sin{\theta_{2}}^{6} < \\ \nonumber 4\sin{\theta_{1}}^{2}\cos{\theta_{1}}^{2}\sin{\theta_{2}}^{2}\cos{\theta_{2}}^{2}.
\label{eq:ineq}
\eea This inequality implies that if $C \ge 1/2$, no background events
of this type will satisfy the requirement on $R$.  if $C \sim 0.4$,
some events can pass, but $M_{T}^{R}$ will reach its allowed maximum,
for fixed $\sqrt{\hat{s}}$, at $M_{T}^{R} \sim \sqrt{\hat{s}}/5$, with
the razor inequality implying that $M_{R} < M_{T}^{R}/C \lesssim
\sqrt{\hat{s}}/2$. Hence for this type of background event to result
in $M_{R} \sim M_{\Delta}$, it must have $\sqrt{\hat{s}} >
2M_{\Delta}$. Therefore, we observe that adding a requirement on $R$
to our event selection will remove most QCD dijet events with
mis-measurements of the type described above. 

Another possibility resulting in non-zero missing transverse momentum
in these background events is that there are additional particles,
whose vectorial sum of transverse momentum is non-zero, that escape
detection. For example, jets resulting from initial state radiation
could remain unseen due to limited detector acceptance, causing a
transverse imbalance in the visible momentum in the event. In order to
understand the effect of this type of background on $M_{R}$ and
$M_{T}^{R}$, we consider the following example. We denote the
vectorial sum of the transverse momentum of particles escaping
detection as $\vec{P}_{T}$. Returning again to the QCD dijet example
described by Eq.~\ref{eq:dijet}, a nonzero value of $\vec{P}_{T}$ will
result in two significant changes to the final state particle
kinematics. Firstly, the missing transverse energy will be non-zero,
with $\vec{M} = \vec{P}_{T}$. Secondly, this missing momentum will
result in the dijet system undergoing an additional transverse boost
when moving from the dijet rest frame to the laboratory frame (any
additional contribution to the longitudinal momentum imbalance in the
event is absorbed into the longitudinal boost, $\beta_{l}$, which
moves the dijets from their CM frame to the laboratory
frame). Specifically, the dijets are moved to a frame traveling at a
velocity $\vec{\beta} = \vec{M}/(\gamma \sqrt{\hat{s}})$, where
$\gamma = (1-|\vec{\beta}|^{2})^{1/2}$ and $\sqrt{\hat{s}}$ is the
dijet invariant mass.  In this case, $M_{R}$ is given by
\begin{equation}
M_{R} = \gamma\sqrt{\hat{s}}\left(1 - \frac{\gamma^{2}(\vec{\beta}\cdot \hat{v})^{2}}{(\hat{v}\cdot \hat{z})^{2}}\right)^{-1/2},
\end{equation}
while $M_{T}^{R}$ can be expressed as
\begin{equation}
M_{T}^{R} \sim \sqrt{\frac{\gamma\beta\hat{s}(\sqrt{(1-(\hat{v}\cdot \hat{z})^{2}}+\gamma\beta)}{2}}.
\end{equation}
We observe that  that for fixed $\sqrt{\hat{s}}$, after applying a requirement on $R$, remaining background events will have $M_{R}$ with an upper bound that goes as $\sqrt{\gamma\beta\hat{s}}$ if the jets have a large transverse component in their rest frame, otherwise as $\gamma\beta\sqrt{\hat{s}}$. Recalling that $\gamma\beta = |\vec{P}_{T}|/\sqrt{\hat{s}}$, we observe that the asymptotic behavior of these upper bounds can be re-expressed as $|\vec{P}_{T}|$ and $(|\vec{P}_{T}|\sqrt{\hat{s}})^{1/2}$, respectively. Hence, we see that in order for these types of background events to populate the $M_{R}$ distribution in the neighborhood of some value of $M_{\Delta}$, the magnitude of the vectorial sum of the transverse momentum of any missing particles needs to be on the order of $M_{\Delta}$. 

In the case of the jets plus missing transverse momentum final state,
this example is not only relevant for the QCD multi-jet background,
but also for the so-called irreducible background
$Z(\nu\nu)+$dijets. Here, $|\vec{P}_{T}| \sim p_{T}^{Z}$, and hence
has an intrinsic scale on the order of $M_{Z}$. The distribution of
$M_{R}$ still falls off exponentially for this background when $M_{Z}
\lesssim M_{R}$.

We observe that $M_{R}$ is potentially a powerful variable for
distinguishing SUSY dijet plus missing transverse momentum events from the
relevant backgrounds when used in conjunction with requirements on the
$R$-frame razor, $R$. In particular, we note that this variable is
robust against effects related to jet mis-measurements and limited
detector phase-space acceptance which often result in spurious missing
transverse momentum.

\section{Generalizing the application of $M_R$ \label{sec:general}}
In Sec.~\ref{sec:RFRAME} we introduced and derived the variable
$M_{R}$ in the context of pair production of two massive particles,
with equal mass, that both decay directly to a mass-less visible
particle and a massive invisible particle. Despite the fact that
$M_{R}$ is motivated by this particular type of example, we find that
it is useful in a more general context.

We return to the example described in Sec.~\ref{sec:RFRAME} but we now
allow for the two massive particles, $G_{1}$ and $G_{2}$, to have
different masses. Alternatively, we can assume that the weakly
interacting particles resulting from the decays of $G_{1}$ and $G_{2}$
have different masses. Using the notation of Sec.~\ref{sec:RFRAME}, we
will assume that each of the two decay chains has a different value
for $M_{\Delta}^{i} =
\frac{M_{G_{i}}^{2}-M_{\chi_{i}}^{2}}{M_{G_{i}}}$, so that
$M_{\Delta}^{2} = M_{\Delta}^{1}(1+\delta) = M_{\Delta}(1+\delta)$.

\begin{figure}[htbp]
\begin{center}
\includegraphics[width=0.4\textwidth]{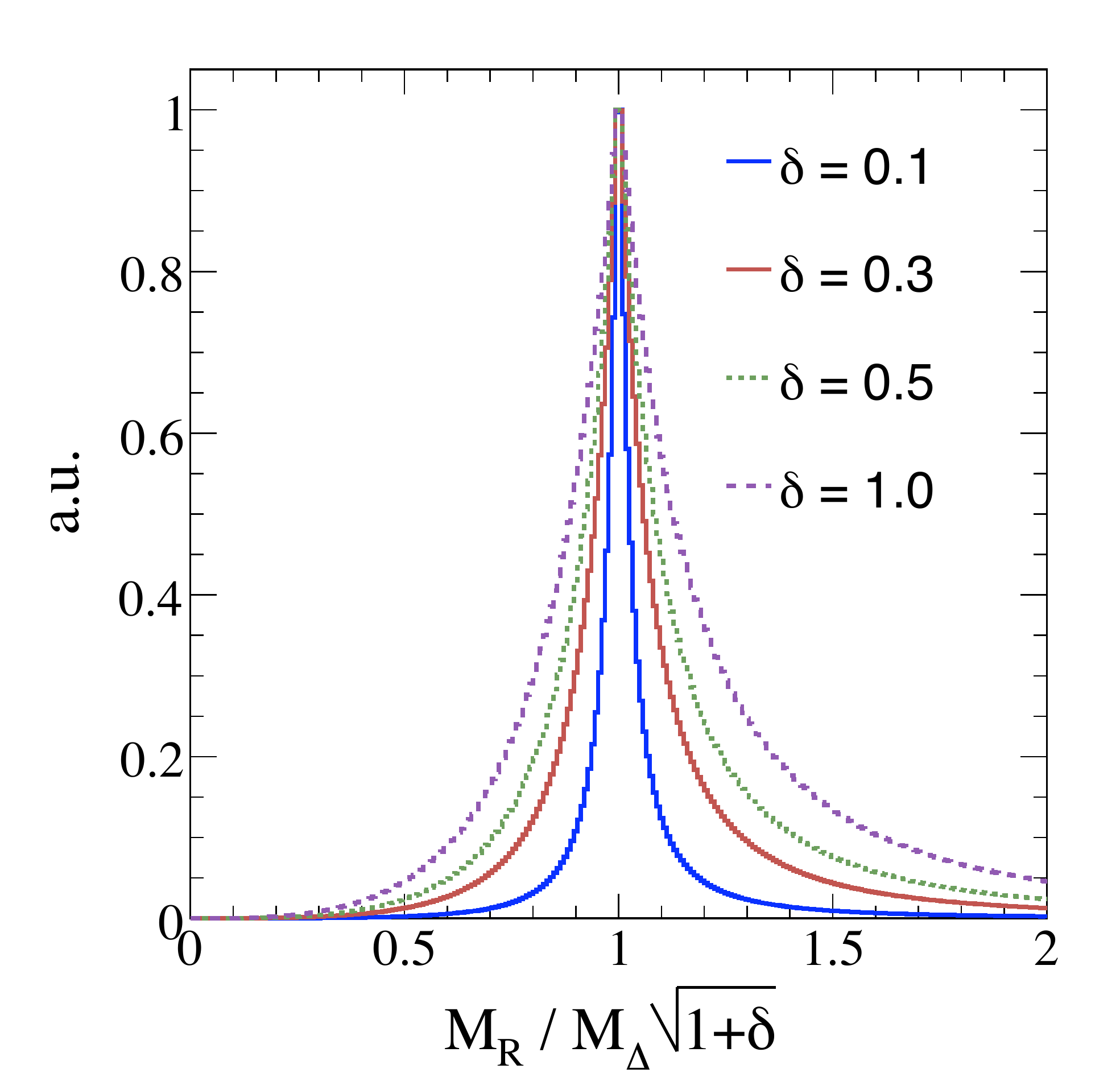}
\caption{Distribution of $M_{R}$, in units of $M_{\Delta}\sqrt{1+\delta}$, for different values of $\delta$. Distributions are normalized such that the maximum value is equal to 1.
  \label{fig:M1M2}}
\end{center}
\end{figure}

Assuming $\gamma_{CM} = 1$, we numerically integrate over the angular degrees of freedom contained in the variables $\hat{u}_{1},\hat{u}_{2}$ to derive the distribution for $M_{R}$, for different values of $\delta$, which is shown in Fig.~\ref{fig:M1M2}. We find that $M_{R}$ peaks precisely at the geometric mean of $M_{\Delta}^{1}$ and $M_{\Delta}^{2}$.

To define $M_{R}$ in cases with more than two visible particles in
the final state, we generalize the two-object to a
multi-object final state by forming two {\it pseudo}-objects, $H_{a}$
and $H_{b}$, with four-vectors $h_{a}$ and $h_{b}$, respectively. Each
pseudo-object's four-momenta is simply the sum of the four-vectors
associated with it.  Each final state object can only be assigned to
one pseudo-object, and each pseudo-object must have at least one
object assigned to it. These associations are determined by minimizing
the quantity $(h_{a})^{2} + (h_{b})^{2}$. Once the two pseudo-objects
are defined, their masses are set to zero with the direction and
magnitude of their momenta unchanged. With this prescription, $M_{R}$
is defined as in Eq.~\ref{eq:MR}, with $q^{l}_{1} = h_{a}$ and
$q^{l}_{2} = h_{b}$.
 
\begin{figure}[htbp]
\begin{center}
\includegraphics[width=0.2385\textwidth]{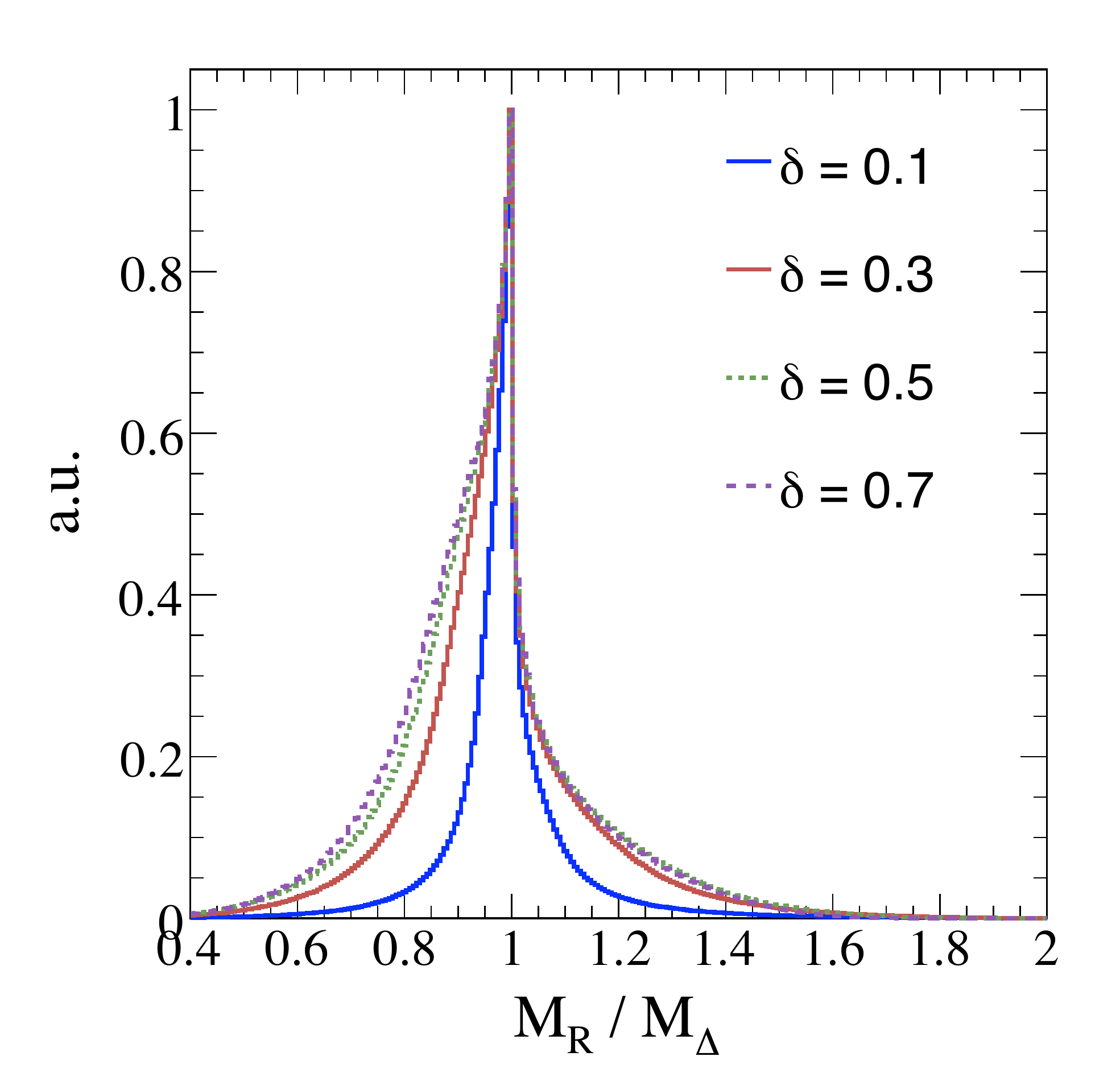}
\includegraphics[width=0.2385\textwidth]{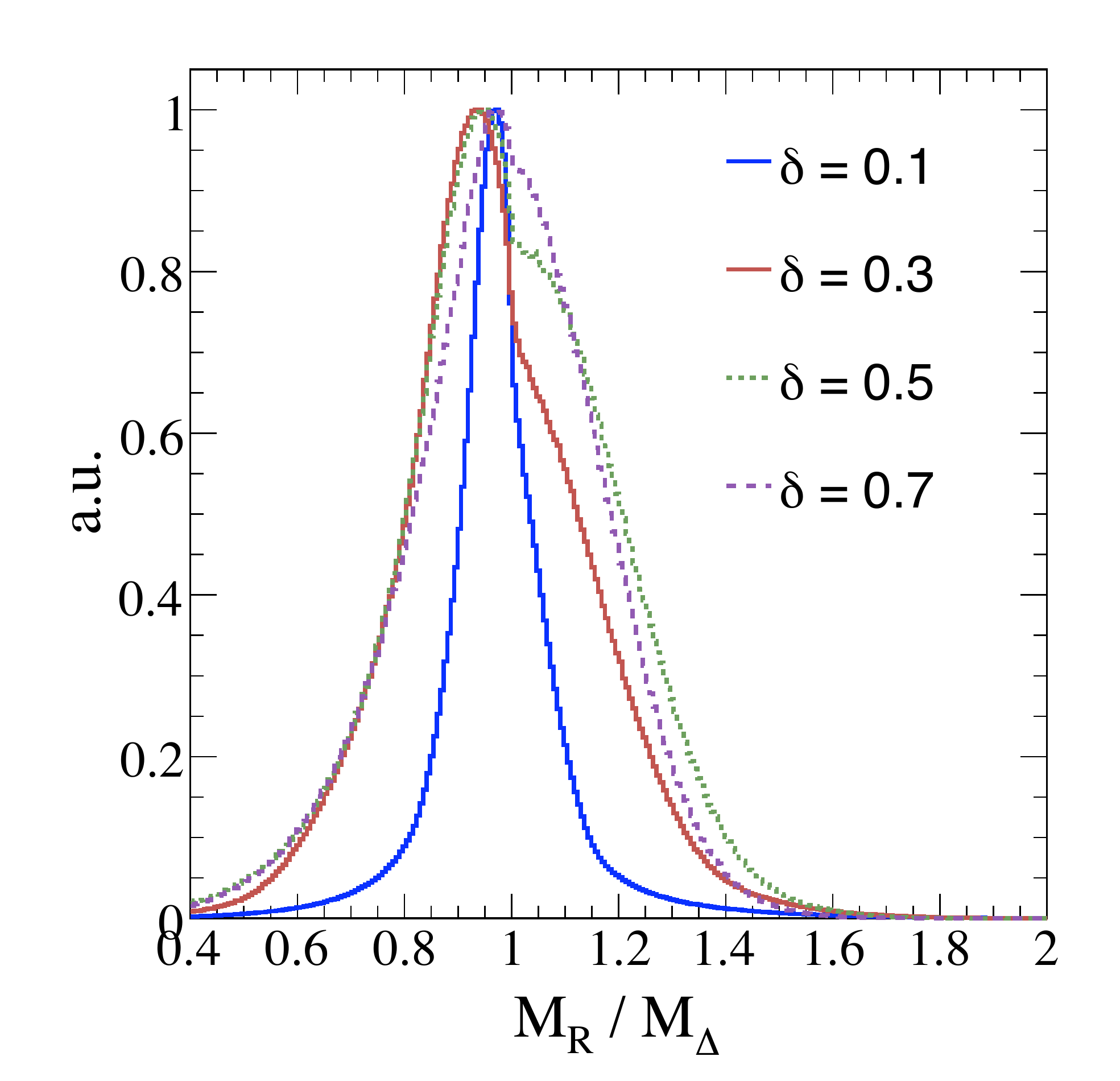}
\caption{Distribution of $M_{R}$ when one (left) or both (right) of the particles $G_{i}$ decays to an intermediate particle $S_{i}$ with mass $M_{S} = M_{G}(1-\delta)$, for different values of $\delta$. Distributions are normalized such that the maximum value is equal to 1
  \label{fig:M1decay}}
\end{center}
\end{figure}

 For example, we assume that the particles $G_{1}$ and $G_{2}$ have the same mass ($M_{G}$), as do $\chi_{1}$ and $\chi_{2}$ ($M_{\chi}$), except now one or both of the particles $G_{i}$ undergoes a two-body decay to a visible particle, $Q_{i+2}$, and another particle, $S_{i}$, with mass $M_{S} = M_{G}(1-\delta)$. The particle $S_{i}$ then decays to another visible particle, $Q_{i}$, and $\chi_{i}$. Numerically integrating over all the decay angles in this scenario (using angular probability distribution functions flat on the unit sphere) with $\gamma_{CM} = 1$, and requiring $R > 0.4$, we derive the distributions for $M_{R}$, assuming one or both of the particles $G_{i}$ decays to an intermediate $S_{i}$,  shown in Fig.~\ref{fig:M1decay}. We find that, in both of these cases, the resulting $M_{R}$ distribution peaks at $M_{\Delta} =  \frac{M_{G}^{2}-M_{\chi}^{2}}{M_{G}}$, regardless of the value of $\delta$ (for the values considered here) and irrespective of whether all of the visible decay products resulting from a particular $G_{i}$ are assigned to the same pseudo-object.
%
\subsection{Example: inclusive search for SUSY\label{sec:SUSY}}

A potential use-case for an analysis incorporating the variables
$M_{R}$ and $R$ is an inclusive search for R-parity conserving SUSY.
Generally, these scenarios can be characterized by the production of
pairs of massive, strongly-interacting super-partners decaying
directly or through cascade decays to SM particles, and LSPs that
escape the detector unseen.

The canonical example described in Sec.~\ref{sec:RFRAME} is the
simplest case of this type of SUSY particle pair production, with the
particles $G_{i}$ representing squarks that each decay directly to a
quark and neutralino. The example introduced earlier in
Sec.~\ref{sec:general} of two heavy particles undergoing two-body decays to intermediate
heavy particles, which subsequently decay to a visible and
weakly-interacting particle, describes the pair-production of gluinos
which decay to a quark and squark, subsequently decaying as in the
canonical example. This cascade decay is often quite prominent when
$M_{\tilde{g}} > M_{\tilde{q}}$.

Hence, for R-parity conserving SUSY scenarios, di-squark production
will result in an $M_{R}$ peak around the scale
$M_{\Delta}^{\tilde{q}\tilde{q}} =
\frac{M_{\tilde{q}}^{2}-M_{\tilde{\chi}}^{2}}{M_{\tilde{q}}}$, with
potentially several different peaks corresponding to the different
squark generations that can only be resolved if the masses are
sufficiently different. If $M_{\tilde{g}} > M_{\tilde{q}}$,
$\tilde{g}\tilde{g}$ and $\tilde{g}\tilde{q}$ production will result
in $M_{R}$ peaks at $M_{\Delta}^{\tilde{g}\tilde{g}} =
\frac{M_{\tilde{g}}^{2}-M_{\tilde{\chi}}^{2}}{M_{\tilde{g}}}$ and
$M_{\Delta}^{\tilde{g}\tilde{q}} =
\sqrt{\left(\frac{M_{\tilde{g}}^{2}-M_{\tilde{\chi}}^{2}}{M_{\tilde{g}}}\right)\left(\frac{M_{\tilde{q}}^{2}-M_{\tilde{\chi}}^{2}}{M_{\tilde{q}}}\right)}$. The
result is $M_{R}$ spectroscopy, with with different peaks
corresponding to the respective mass differences between the massive
SUSY particles produced first in the hard scattering process and the
LSP, and the geometric means of these mass differences when two SUSY
particles are produced with different masses.

If one approaches the jets + missing transverse energy + $X$ final
state in an inclusive way, the dominant backgrounds will be QCD
multi-jets, $t\bar{t}+$jets and $V+$jets, where $V$ is a $W$ or $Z$
vector boson decaying to leptons and/or neutrinos. The QCD
contribution to this background will be largely marginalized by the
use of a cut on the razor, $R$, as described in
Sec.~\ref{sec:razor}. The $M_{R}$ distribution of QCD events passing
this requirement will fall roughly exponentially in $M_{R}$, with a
slope largely determined by the value of the cut on $R$, but not
exceeding the slope of the $\sqrt{\hat{s}}$ for these QCD
processes. The remaining backgrounds identified above must have a
large transverse momentum imbalance in the event in order to pass the
requirement on $R$, implying that their contribution to the
distribution in $M_{R}$ will be comprised of events with final state
neutrinos or leptons that are not explicitly identified as such. As
was discussed in Sec.~\ref{sec:razor} with the example of
$Z(\nu\nu)$+jets, the $M_{R}$ distributions for these processes will
also fall roughly exponentially, with slope determined predominantly
by each processes' respective scale, $M_{Z}$, $M_{W}$ and $M_{t}$.

As a result, if the various values of $M_{\Delta}$ that characterizes a particular SUSY scenario are sufficiently higher than $M_{Z}$, $M_{W}$ and $M_{t}$, the signal events will appear as a wide peak(s) on top of falling exponential backgrounds, potentially a  striking signature depending on the relative production rates of the SUSY and background processes.

\subsection{Example: $H\to W(\ell\nu)W(\ell\nu)$\label{sec:HIGGS}}

We have, so far, only discussed the  canonical scenario described in Sec.~\ref{sec:RFRAME} in the context of cases where $\gamma_{CM}$ is near one, corresponding to the pair-production of massive particles near threshold. The same example applies to the process of a Higgs boson decaying to two $W$ bosons, which subsequently decay leptonically. Now, $\gamma_{CM} = M_{H}/2M_{W}$.

\begin{figure}[htbp]
\begin{center}
\includegraphics[width=0.4\textwidth]{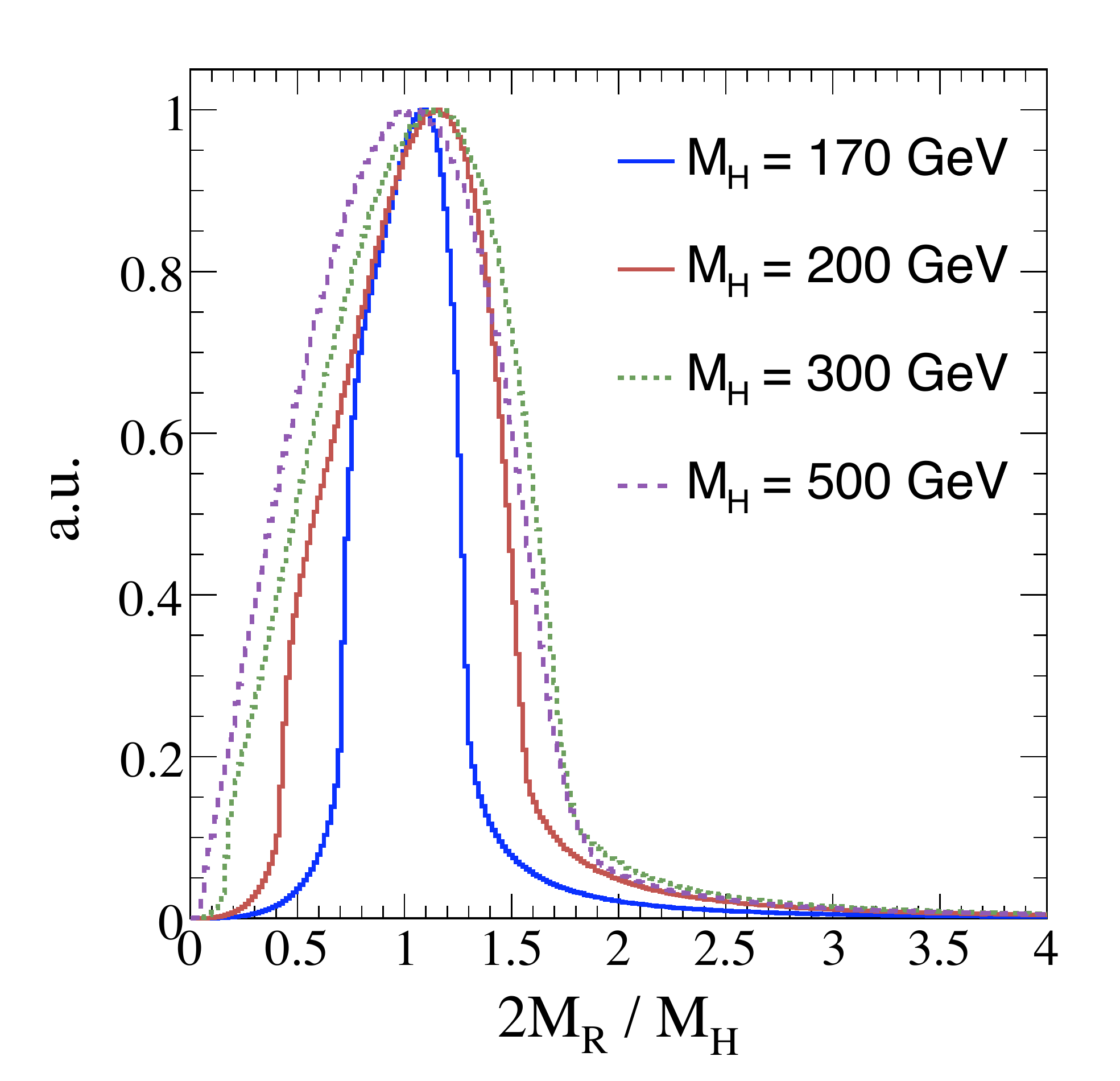}
\caption{Distribution of $M_{R}$ for the di-lepton final state in $H\to WW$ as a function of $M_{H}$, where we have made the approximations  $p_{T}^{H} = 0$ and neglected spin-correlations. 
  \label{fig:MH}}
\end{center}
\end{figure}

As we saw in Sec.~\ref{sec:gCM}, the peak value of $M_{R}$ will scale as $\gamma_{CM} M_{\Delta}$ which, in this case, implies that $M_{R}$ will peak roughly at $M_{H}/2$. Fig.~\ref{fig:MH} shows the distribution of $M_{R}$ for different values $M_{H}$, with $p_{T}^{H} = 0$.

\begin{figure}[htbp]
\begin{center}
\includegraphics[width=0.2385\textwidth]{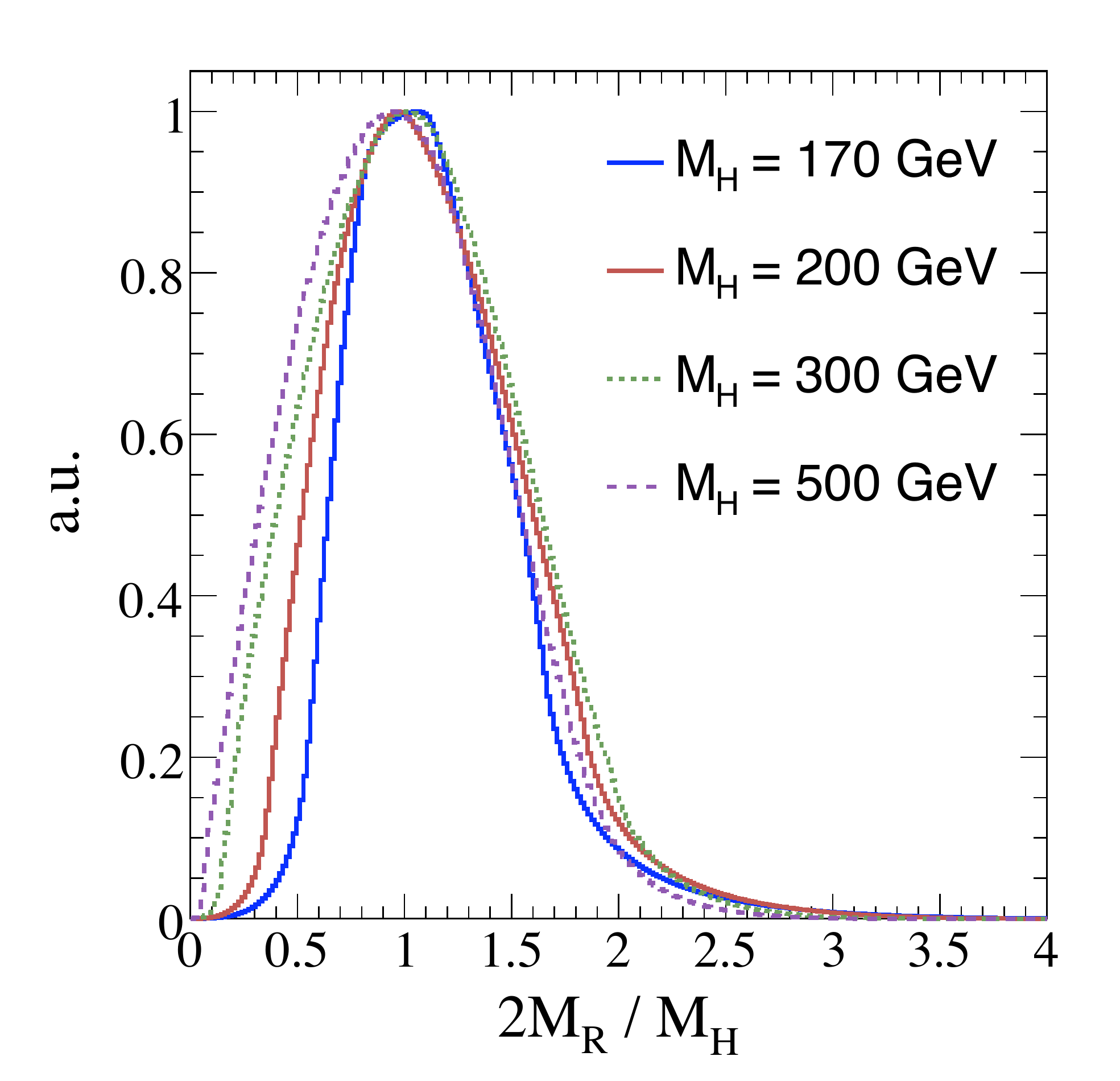}
\includegraphics[width=0.2385\textwidth]{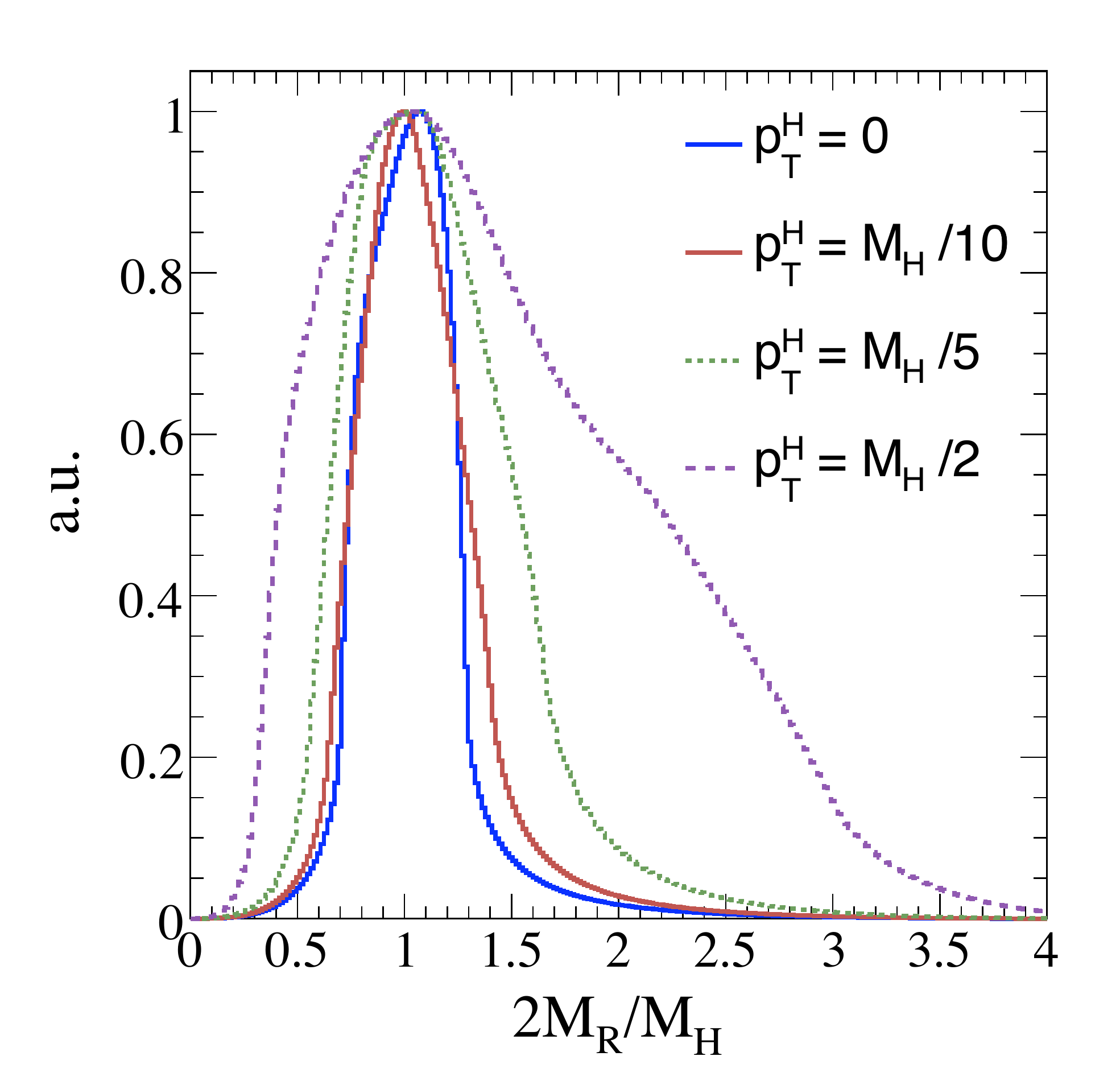}
\caption{Distribution of $M_{R}$ when: (left) $p_{T}^{H} = M_{R}/5$. (right) $M_{H} = 170$ GeV. Spin-correlations are neglected in both cases. 
  \label{fig:MH_pt}}
\end{center}
\end{figure}

In practice we know that $p_{T}^{H} \ne 0$, and, in fact, it scales
with $M_{H}$. In Fig.~\ref{fig:MH_pt} we illustrate the effects of
non-zero $p_{T}^{H}$. We show in Fig.~\ref{fig:MH_pt} (left) the
distribution of $M_{R}$ for different values of $M_{H}$, with
$p_{T}^{H} = M_{H}/5$. In Fig.~\ref{fig:MH_pt} (right) we illustrate
the situation where $M_{H} = 170$ GeV for different values of
$p_{T}^{H}$.

We observe that $2M_{R}$  peaks around the true value of $M_{H}$. This peak is naturally quite wide, such that the lepton reconstruction resolution will have a negligible effect on its shape. An analysis using $M_{R}$ will also benefit from a selection including a requirement on $R$, which should marginalize backgrounds where one or both leptons results from weakly decaying hadrons. Other backgrounds, such as $Z(\ell\ell)$ and $t\bar{t}$ production, as was the case in the jets + missing transverse energy final state described in Sec.~\ref{sec:razor}, will fall roughly exponentially in $M_{R}$ once $M_{R}$ exceeds their respective mass scales.

\section{$M_{R^{*}}$  \label{sec:RSTAR}}

In Sec.~\ref{sec:RFRAME}, we described how to move from the laboratory frame to the $R$-frame via a longitudinal boost, $\beta_{R}$. With the assumptions described in that section, particularly that $\gamma_{CM} = 1$, we found that the variable $M_{R}$ is equal to $M_{\Delta}$. In subsequent sections we discussed the properties of the variable $M_{R}$ when some of these assumptions no longer apply and for deviations from the simple scenario described in Sec.~\ref{sec:RFRAME}. We have shown that the useful properties of the variable $M_{R}$ are robust against the variations we have considered,
but there is an important caveat.

Namely, when $\gamma_{CM}$ deviates from one, there are situations when
$|\beta_{R}| \ge 1$, such that $\beta_{R}$ no longer describes a
physical boost and the $R$-frame is ill-defined, as is the variable
$M_{R}$. In this section we will describe a set of
reference frames, which we will denote the $R^{*}$-frames, which are always well-defined, and variables $R^{*}$ and $M_{R^{*}}$.

\begin{figure}[htbp]
\begin{center}
\includegraphics[width=0.49\textwidth]{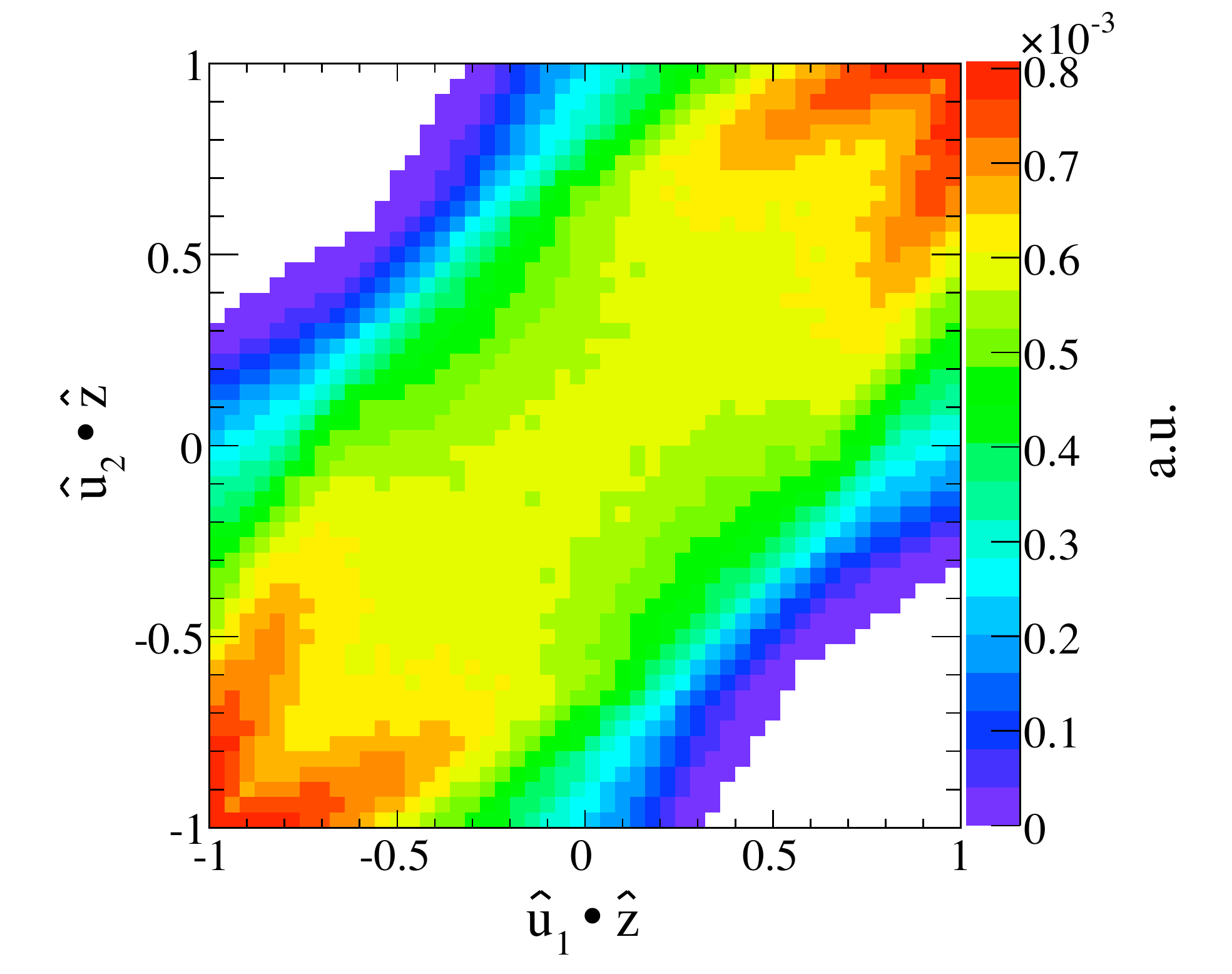}
\caption{Correlation between $\hat{u}_{1}\cdot \hat{z}$ and $\hat{u}_{2}\cdot \hat{z}$ for events with $\gamma_{CM} = 1.1$ and $|\beta_{R}| \ge 1$. Distribution is normalized to unit volume.
  \label{fig:ALT_config_z}}
\end{center}
\end{figure}

Firstly, we identify the  kinematical characteristics that are associated with events with $|\beta_{R}| \ge 1$. Using the notation of Sec.~\ref{sec:RFRAME}, we again consider the pair-production of massive particles resulting in two visible particles, $Q_{1}$ and $Q_{2}$, along with missing transverse energy $\vec{M}$. Setting $\gamma_{CM} = 1.1$, we scan over values for the unit vectors $\hat{u}_{1}$, $\hat{u}_{2}$ and $\hat{\beta}_{CM}$, noting for which values and with what frequency we find $|\beta_{R}| \ge 1$. In Fig.~\ref{fig:ALT_config_z} we show the correlation between the normalized $z$-components of momenta of $Q_{1}$ and $Q_{2}$ in the rest frames of their respective parents $G_{i}$ for events where the $R$-frame is ill-defined. We find, as perhaps one could infer from the expression of $\beta_{R} = \frac{q^{l}_{10}-q^{l}_{20}}{q^{l}_{1z}-q^{l}_{2z}}$, that these longitudinal momentum components tend to be equal in both direction and magnitude. In fact, as $\gamma_{CM}$ tends toward one, the distribution shown in Fig.~\ref{fig:ALT_config_z} tends toward a discrete line along the $\hat{u}_{1}\cdot \hat{z} = \hat{u}_{2}\cdot \hat{z}$ diagonal.

\begin{figure}[htbp]
\begin{center}
\includegraphics[width=0.49\textwidth]{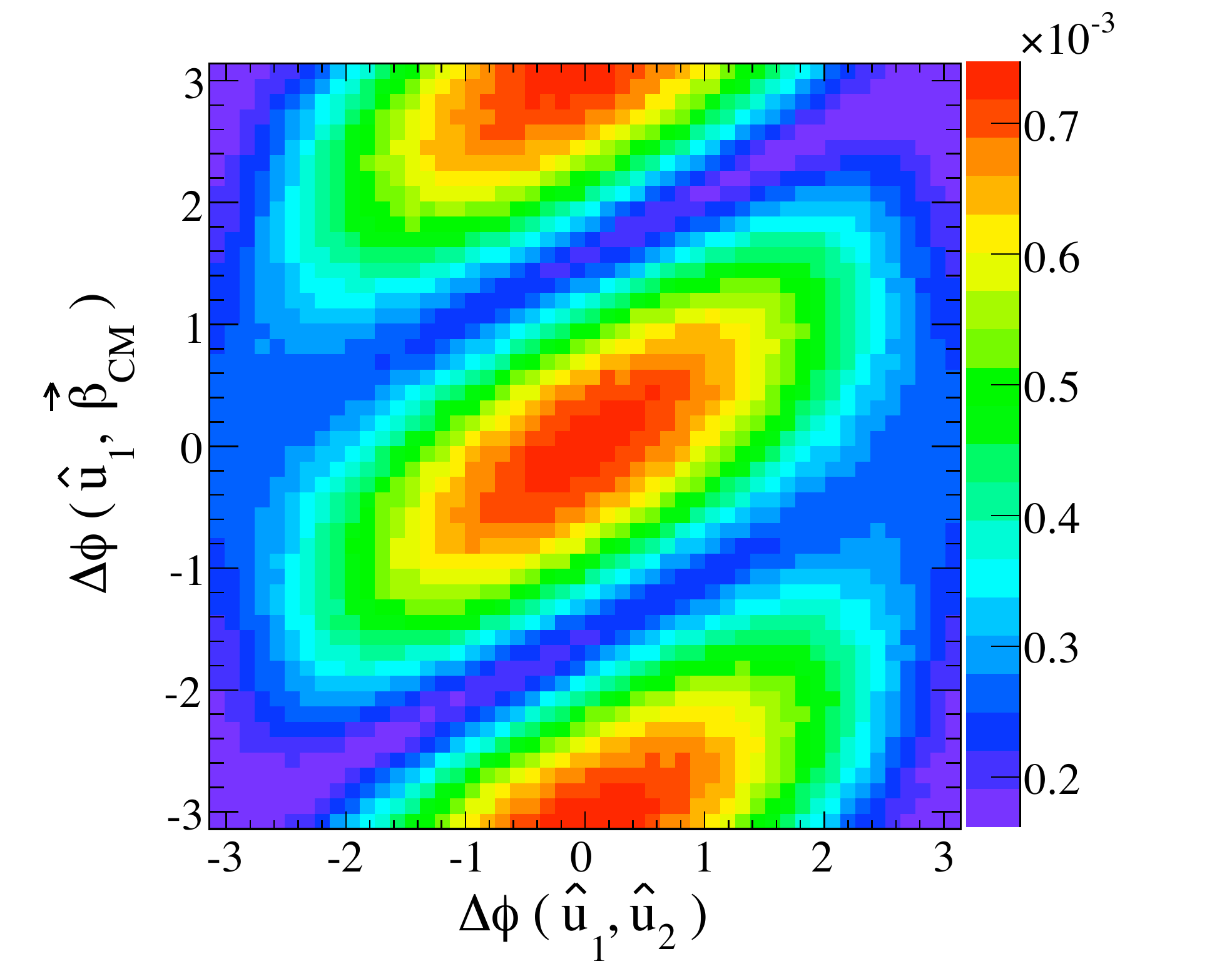}
\caption{Correlation between $\Delta\phi(\hat{u}_{1},\hat{u}_{2})$ and $\Delta\phi(\hat{u}_{1},\vec{\beta}_{CM})$ for events with $\gamma_{CM} = 1.1$ and $|\beta_{R}| \ge 1$. Distribution is normalized to unit volume.
  \label{fig:ALT_config_dphi}}
\end{center}
\end{figure}

In Fig.~\ref{fig:ALT_config_dphi} we show the correlation between the difference in azimuthal angles between the momenta of $Q_{1}$ and $Q_{2}$ and between the momenta of $Q_{1}$ and $\vec{\beta}_{CM}$. We find that events with $|\beta_{R}| \ge 1$ tend to have $\hat{u}_{1}$ and $\hat{u}_{2}$ pointing in the same direction in the transverse plane, with $\vec{\beta}_{CM}$ pointing in either the same or opposite direction. 

These observations indicate that the cases where the $R$-frame is ill-defined result from the neglecting of the transverse component of $\vec{\beta}_{CM}$ in the approximations made in the derivation of $M_{R*}$. Here, we describe the derivation of variables that take into account this transverse component.

We consider a longitudinal boost to the four-vectors $q^{l}_{1}$ and $q^{l}_{2}$, associated with a velocity $\beta_{L^{*}}$. Subsequently, we consider a transverse boost, with velocity $\vec{\beta}_{T}^{R^{*}}$, that is applied in {\it opposite} directions to the four-vectors associated with $Q_{1}$ and $Q_{2}$, taking them to different reference frames.  We denote these two reference frames as $R^{*}$-frames, with the additional requirement that, in each of their respective $R^{*}$-frames, the momenta of $Q_{1}$ and $Q_{2}$ must be equal in magnitude.  The series of Lorentz boosts taking $Q_{1}$ and $Q_{2}$ from the laboratory frame to their respective $R^{*}$-frames can be summarized as:
\bea
q_1^{l} \xrightarrow{\beta_{L^{*}}} q^{b}_1 \xrightarrow{\vec{\beta}_{T}^{R^{*}}} q^{R^{*}}_1
\nonumber\\
q_2^{l} \xrightarrow{\beta_{L^{*}}} q^{b}_2 \xrightarrow{-\vec{\beta}_{T}^{R^{*}}} q^{R^{*}}_2
\eea

The constraint that $Q_{1}$ and $Q_{2}$ have the same energy in the respective $R^{*}$ frames can be re-expressed as a constraint equation on the variables $\beta_{L^{*}}$ and $\vec{\beta}_{T}^{R^{*}}$:
\begin{equation}
\gamma_{L^{*}}(q_{10}^{l}-q_{20}^{l})-\gamma_{L^{*}}\beta_{L^{*}}(q_{1z}^{l}-q_{2z}^{l}) = \vec{\beta}_{T}^{R^{*}}\cdot~(\vec{q}_{1T}^{l}+\vec{q}_{2T}^{l})
\end{equation}
which can be used to solve for $|\vec{\beta}_{T}^{R^{*}}| \equiv \beta_{T}^{R^{*}}$ in terms of $\hat{\beta}_{T}^{R^{*}}$ and $\beta_{R^{*}}$:
\begin{equation}
\beta_{T}^{R^{*}} = \frac{\gamma_{L^{*}}(q_{10}^{l}-q_{20}^{l})-\gamma_{L^{*}}\beta_{L^{*}}(q_{1z}^{l}-q_{2z}^{l})}{\hat{\beta}_{T}^{R^{*}}\cdot~(\vec{q}_{1T}^{l}+\vec{q}_{2T}^{l})} .
\end{equation}

Just as we did in the $R$-frame, we will define the $R^{*}$-frame mass, $M_{R^{*}}$, as two times the magnitude of the momentum of $Q_{1}$ in it's respective $R^{*}$-frame. $M_{R^{*}}$ can be expressed as:
\bea
M_{R^{*}} \equiv 2|\vec{q}_{1}^{R^{*}}| = 2|\vec{q}_{2}^{R^{*}}| = 
\nonumber\\
\frac{2\gamma_{L^{*}}\hat{\beta}_T^{R^{*}}\cdot \left[(q_{10}^{l}\vec{q}_{2T}^{l} + q_{20}^{l}\vec{q}_{1T}^{l})-\beta_{L^{*}}(q_{1z}^{l}\vec{q}_{2T}^{l} + q_{2z}^{l}\vec{q}_{1T}^{l})\right]   }{\sqrt{|\hat{\beta}_{T}^{R^{*}}\cdot (\vec{q}_{1T}^{l}+\vec{q}_{2T}^{l})|^{2}-\gamma_{L^{*}}\left[ q_{10}^{l}-q_{20}^{l}-\beta_{L^{*}}(q_{1z}^{l}-q_{2z}^{l})\right]^{2}}}\nonumber\\ .
\eea

In order to calculate $M_{R^{*}}$, we must specify values of $\beta_{L^{*}}$ and  $\hat{\beta}_{T}^{R^{*}}$. Motivated by the event configurations which lead to $|\beta_{R}| \ge 1$, in particular those described in Fig.~\ref{fig:ALT_config_dphi}, we choose a value of $\hat{\beta}_{T}^{R^{*}}$ (an angle in the azimuthal plane) which maximizes the quantity $|\hat{\beta}_{T}^{R^{*}}\cdot (\vec{q}_{1T}^{l}+\vec{q}_{2T}^{l})|$. With this choice, $\hat{\beta}_{T}^{R^{*}}$ can be expressed as:
\begin{equation}
\hat{\beta}_{T}^{R^{*}} = \frac{\vec{q}_{1T}^{l}+\vec{q}_{2T}^{l}}{|\vec{q}_{1T}^{l}+\vec{q}_{2T}^{l}|}
\end{equation}
The value of $\beta_{L^{*}}$ is determined by requiring that the condition $\frac{\partial M_{R^{*}}}{\partial \beta_{L^{*}}} = 0$ is satisfied. This choice results in
\begin{equation}
\beta_{L^{*}} = \frac{q_{1z}^{l}+q_{2z}^{l}}{q_{10}^{l}+q_{20}^{l}}
\end{equation}
With each unknown quantity now specified, $M_{R^{*}}$ can be expressed, event-by-event, as
\begin{equation}
M_{R^{*}} = \sqrt{(q_{10}^{l}+q_{20}^{l})^{2}-(q_{1z}^{l}+q_{2z}^{l})^{2}-\frac{(|\vec{q}^{l}_{1T}|^{2}-|\vec{q}^{l}_{2T}|^{2})^{2}}{|\vec{q}^{l}_{1T}+\vec{q}^{l}_{2T}|^{2}}}
\end{equation}

Another interesting quantity is $\gamma_{R^{*}} = (1-|\vec{\beta}_{T}^{R^{*}}|^{2})^{-1/2}$, which can be expressed in terms of lab frame observables as
\bea
\gamma_{R^{*}} = \sqrt{\frac{(q_{10}^{l}+q_{20}^{l})^{2}-(q_{1z}^{l}+q_{2z}^{l})^{2}}{(q_{10}^{l}+q_{20}^{l})^{2}-(q_{1z}^{l}+q_{2z}^{l})^{2}-\frac{(|\vec{q}^{l}_{1T}|^{2}-|\vec{q}^{l}_{2T}|^{2})^{2}}{|\vec{q}^{l}_{1T}+\vec{q}^{l}_{2T}|^{2}}}}
\eea
As is the case for $M_{R}$, $M_{R^{*}}$ is invariant under longitudinal boosts, as is $\gamma_{R^{*}}$. 

Similarly as for the $R$-frame, we define the $R^{*}$-frame razor, $R^{*}$, as the ratio of $M_{T}^{R}$ and $M_{R^{*}}$, with $M_{T}^{R}$ given by Eq.~\ref{eq:MT_R}.

\begin{figure}[htbp]
\begin{center}
\includegraphics[width=0.2385\textwidth]{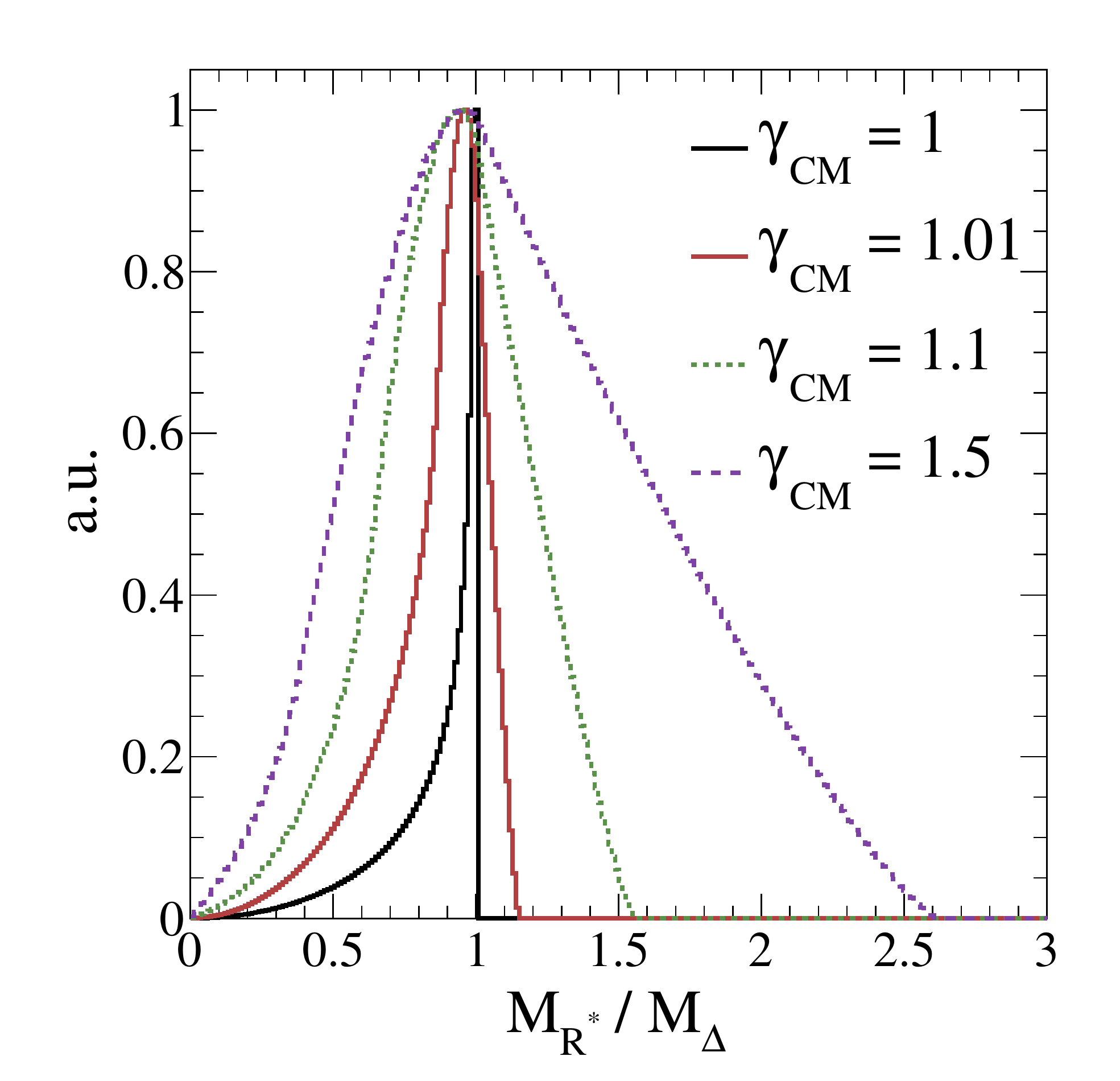}
\includegraphics[width=0.2385\textwidth]{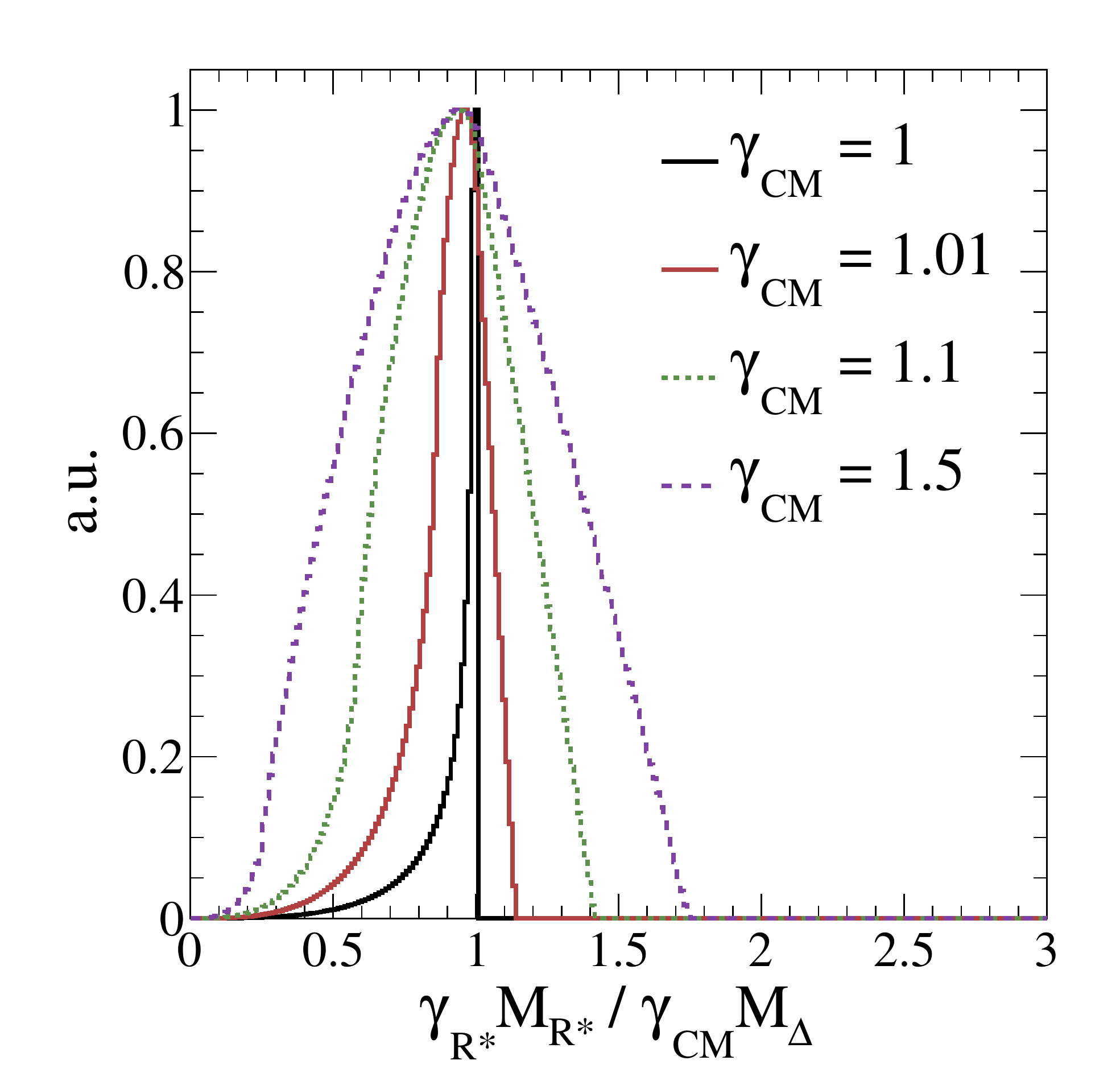}
\caption{Distribution of $M_{R^{*}}$ (left) and $\gamma_{R^{*}}M_{R^{*}}$ (right) for different values of $\gamma_{CM}$. Distributions are normalized such that their maximum value is equal to one.
  \label{fig:MRstar_can}}
\end{center}
\end{figure}

To understand how the distribution of $M_{R^{*}}$ changes with $\gamma_{CM}$, we numerically integrate over all the decay angles, assuming their distributions are flat on the unit sphere. The resulting $M_{R^{*}}$ and $\gamma_{R^{*}}M_{R^{*}}$ distributions are shown in Fig.~\ref{fig:MRstar_can}. We find that the peak value of the $M_{R^{*}}$ distribution is at approximately $M_{\Delta}$, regardless of $\gamma_{CM}$, while $\gamma_{R^{*}}M_{R^{*}}$ peaks at $\gamma_{CM}M_{\Delta}$. 

Comparing Fig.~\ref{fig:MRstar_can} and Fig.~~\ref{fig:MR_can}, we see that the peak position of the $\gamma_{R^{*}}M_{R^{*}}$ distribution scales like the peak of the $M_{R}$ distribution. $M_{R}$ is a variable most useful for treating the case $\gamma_{CM} =1$ which, in practice, is kinematically forbidden. The quantity $\gamma_{R^{*}}M_{R^{*}}$ reproduces the same peaking behavior, without ill-defined configurations and better resolution on the quantity $\gamma_{CM}M_{\Delta}$.

\begin{figure}[htbp]
\begin{center}
\includegraphics[width=0.2385\textwidth]{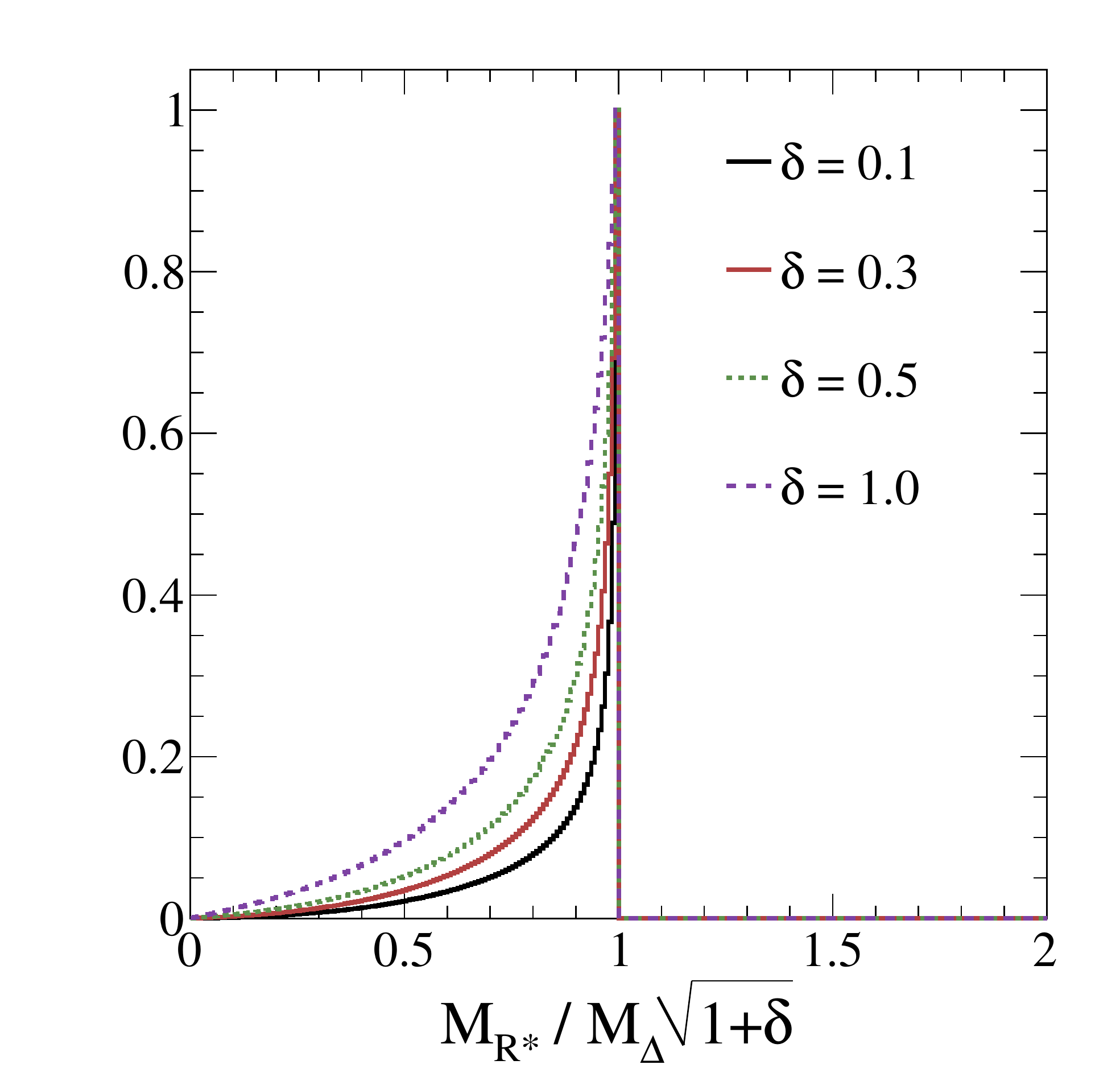}
\includegraphics[width=0.2385\textwidth]{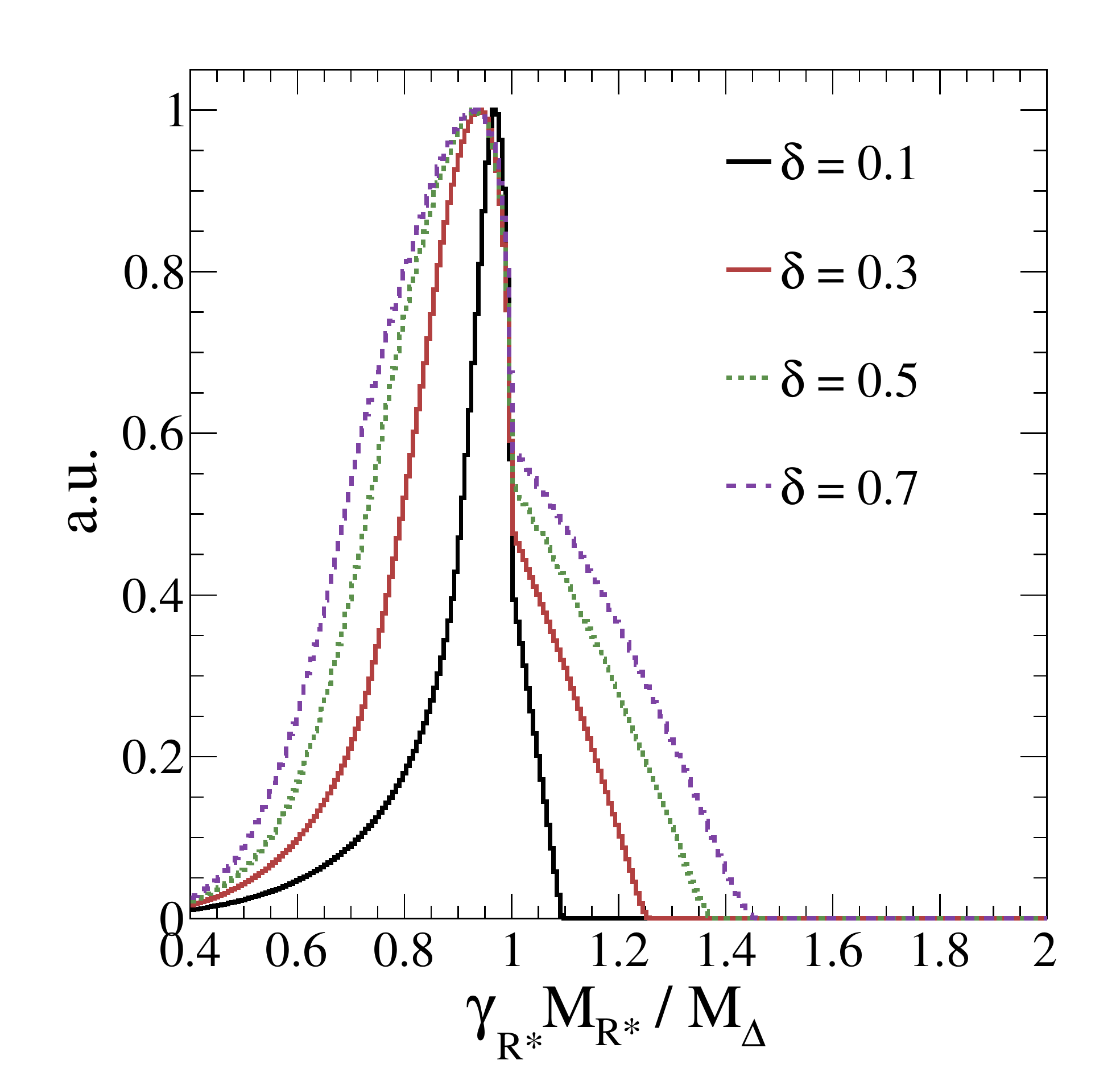}
\caption{(left) Distribution of $M_{R^{*}}$, in units of $M_{\Delta}\sqrt{1+\delta}$, for different  values of $\delta$. (right) Distribution of $\gamma_{R^{*}}M_{R^{*}}$ when one of the particles $G_{i}$ decays to an intermediate particle $S_{i}$ with mass $M_{S} = M_{G}(1-\delta)$, for different values of $\delta$. Distributions are normalized such that their maximum value is equal to one.
  \label{fig:MRstar_example}}
\end{center}
\end{figure}

In fact, the variables $M_{R^{*}}$, $\gamma_{R^{*}}M_{R^{*}}$ and $M_{R}$ share many properties. We consider two of the examples from Sec.~\ref{sec:general}, now in the context of  $M_{R^{*}}$ and $\gamma_{R^{*}}$. The first scenario is of two massive particles, $G_{1}$ and $G_{2}$, with different masses decaying each to a visible particle and potentially massive weakly interacting particle, such that $M_{\Delta}^{2} = M_{\Delta}^{1}(1+\delta) = M_{\Delta}(1+\delta)$. Assuming $\gamma_{CM} = 1$, and numerically integrating over the angular degrees of freedom assuming scalar decays, we calculate $M_{R^{*}}$ as a function of $\delta$, with the resulting distributions shown in Fig.~\ref{fig:MRstar_example} (left). We observe that $M_{R^{*}}$, like $M_{R}$, has a peak whose position scales with $\sqrt{1+\delta}$. 

The second example from Sec.~\ref{sec:general} involves two particles $G_{1}$ and $G_{2}$, with the same mass. $G_{1}$ undergoes a two-body decay to a visible particle, $Q_{3}$, and another particle, $S_{1}$, with mass $M_{S} = M_{G}(1-\delta)$. The particles $G_{2}$ and $S_{1}$ then each decay to a weakly interacting particle and a visible particle, where the mass of the weakly interacting particles is $M_{\chi}$. The numerically integrated $\gamma_{R^{*}}M_{R^{*}}$ distributions, for different values of $\delta$, are shown in Fig.~\ref{fig:MRstar_example} (right). We observe that, like $M_{R}$, the quantity $\gamma_{R^{*}}M_{R^{*}}$ peaks at $M_{\Delta} = \frac{M_{G}^{2}-M_{\chi}^{2}}{M_{G}}$, regardless of the value of $\delta$.

The interplay between $R^{*}$ and $M_{R^{*}}(\gamma_{R^{*}}M_{R^{*}})$ is qualitatively the same as between $R$ and $M_{R}$ for di-jet backgrounds, as described in Sec.~\ref{sec:razor}. A requirement on $R^{*}$ suppresses contributions from mis-measured di-jet events significantly, and the quantity $\gamma_{R^{*}}M_{R^{*}}$ peaks at $\sqrt{\hat{s}}$. The asymptotic behavior of $M_{R^{*}}$ is similar for backgrounds with an escaping (weakly interacting or outside acceptance), high transverse momentum, particle or system of particles.  We note that  $M_{R^{*}}(\gamma_{R^{*}}M_{R^{*}})$ can be used similarly to $M_{R}$ in the context of an inclusive SUSY search, as described in Sec.~\ref{sec:SUSY} or in the search for $H \to WW \to \ell\nu\ell\nu$, as described in Sec.~\ref{sec:HIGGS}. 

\begin{figure}[htbp]
\begin{center}
\includegraphics[width=0.2385\textwidth]{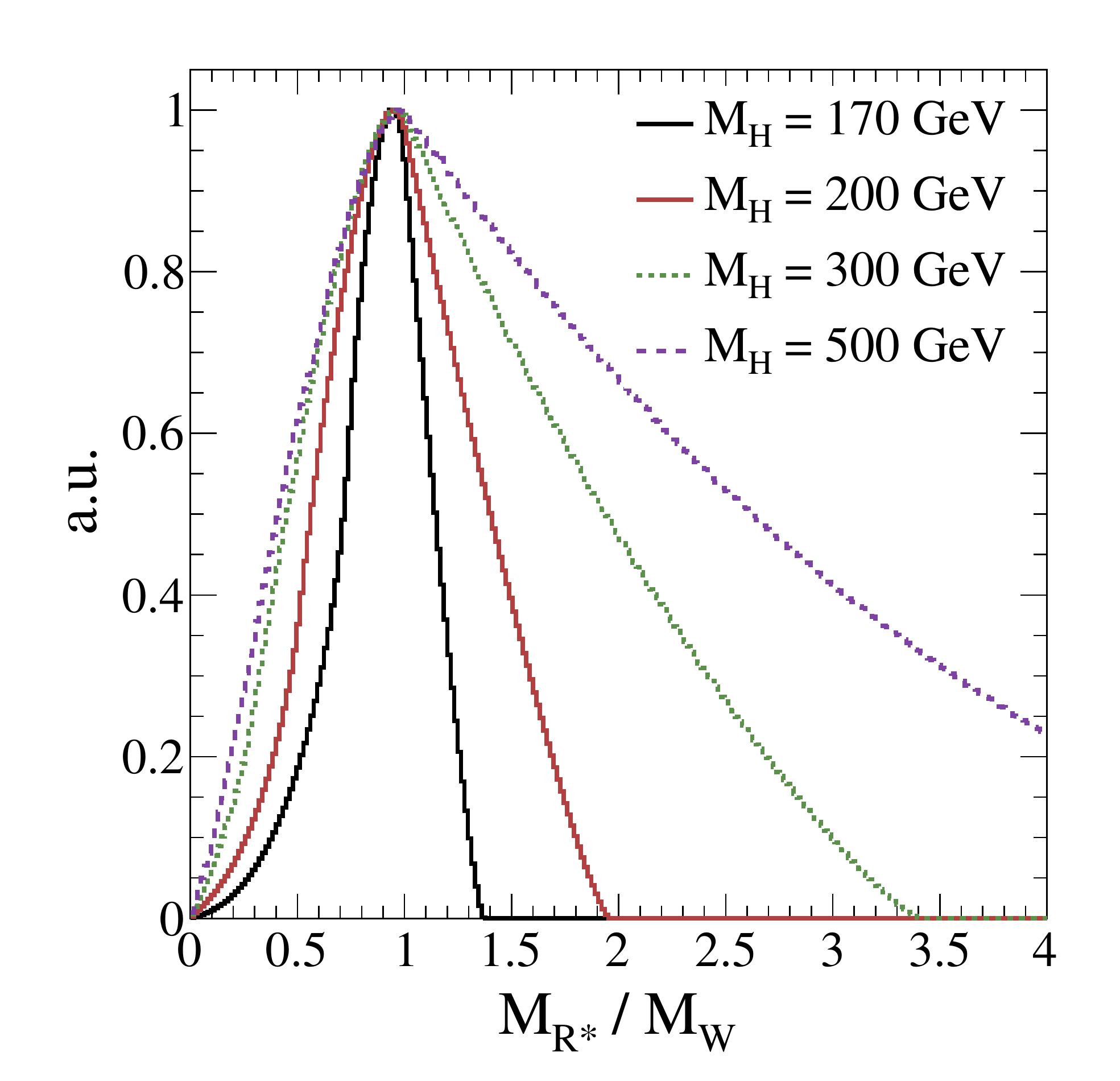}
\includegraphics[width=0.2385\textwidth]{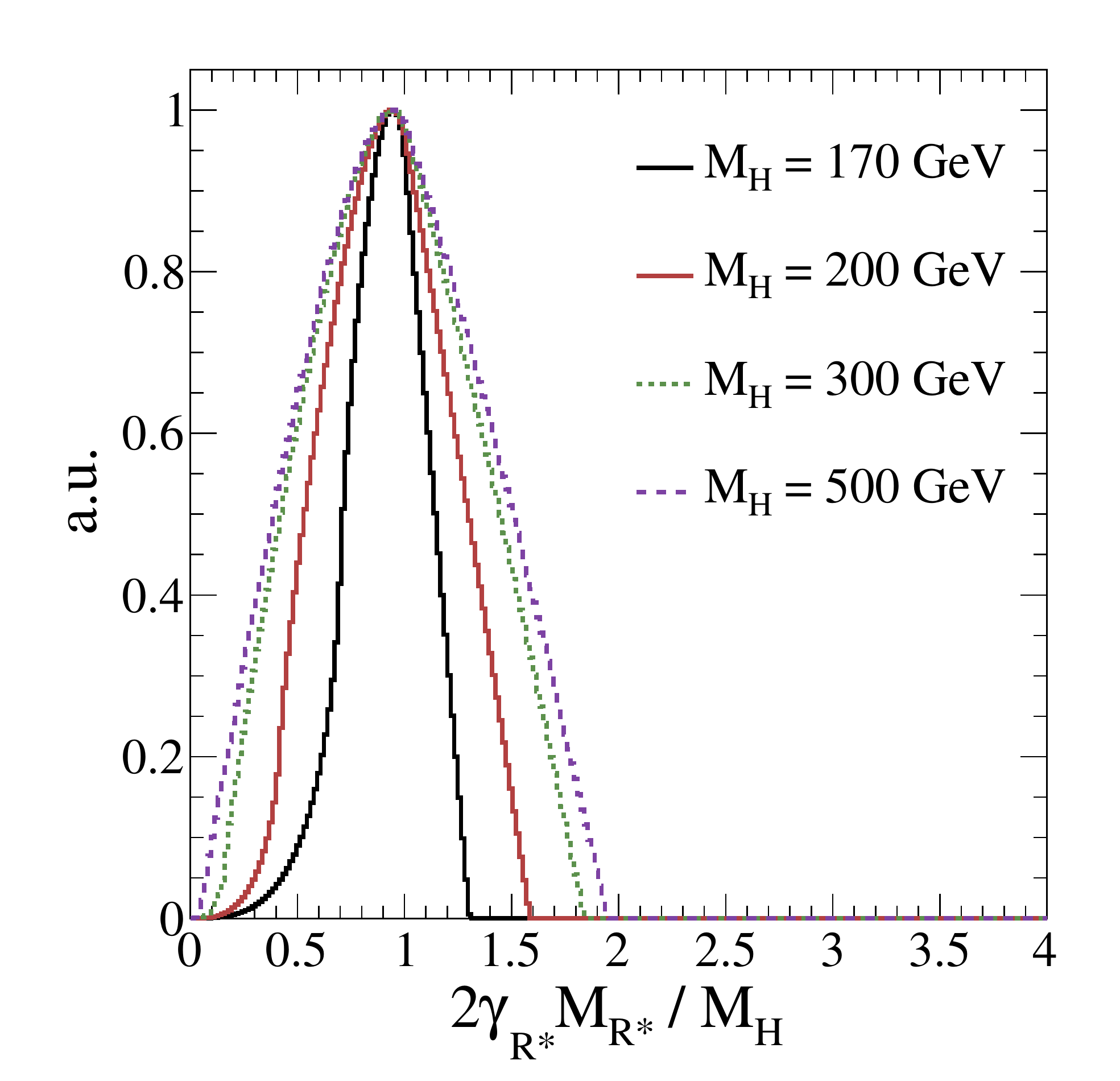}
\caption{(left) Distributions of $M_{R^{*}}$ and (right) $\gamma_{R^{*}}M_{R^{*}}$ for $H\to WW \to \ell\nu\ell\nu$ events.. We make the approximation that $p_{T}^{H} = 0$ and spin correlations are neglected.
  \label{fig:MHstar}}
\end{center}
\end{figure}

In the latter case, we  have an additional piece of information when using $M_{R^{*}}(\gamma_{R^{*}}M_{R^{*}})$ rather than $M_{R}$ in $\gamma^{R^{*}}$. We find that for $H \to WW \to \ell\nu\ell\nu$ decays, $M_{R^{*}}$ will peak at $M_{W}$ while $\gamma^{R^{*}}M_{R^{*}}$ will peak at $M_{H}/2$. These two observations are demonstrated in Fig.~\ref{fig:MHstar}.

\subsection{Outlook}

We introduce a set of variables $M_{R}$ and $M_{R^{*}}$ designed to study the characteristics of processes involving the pair-production of massive particles that each decay directly or through a cascade of decays to SM particles and weakly-interacting particles escaping detection.  We also describe the dimension-less variables $R$ and $R^{*}$ which, used in conjunction with $M_{R}$ and $M_{R^{*}}$, provide the means to select these processes in the presence of large backgrounds, for a variety of final states. We have described how these variables can be used to discover and characterize R-parity conserving SUSY scenarios and SM Higgs boson decays to leptonically decaying $W$-bosons.
\\
\\

 \subsection*{Acknowledgments}
The author would like to thank the CMS collaboration for useful discussions and the Planck 2010 organizers for hosting a talk on the subject. 

This work is supported in part by the U.S. Dept. of Energy under contact DE-FG02-92-ER40701.



\end{document}